\documentclass[prb,twocolumn, amsmath,amssymb]{revtex4}

\setcitestyle{numbers,square}
\usepackage{amsmath,amssymb}
\usepackage{tabularx}
\usepackage{graphicx}
\usepackage{bmpsize}
\usepackage{bm}
\usepackage{color}
\usepackage{amssymb}
\usepackage[version=3]{mhchem}
\usepackage{newtxmath}
\DeclareMathAlphabet{\mathpzc}{OT1}{pzc}{m}{it}

\def\GVec#1{\mbox{\boldmath $#1$}}

\begin{document}
\title{
Multiorbital edge and corner states in black phosphorene
}
\author{Masaru Hitomi}
\affiliation{Department of Physics, Osaka University, Toyonaka, Osaka 560-0043, Japan}
\author{Takuto Kawakami}
\affiliation{Department of Physics, Osaka University, Toyonaka, Osaka 560-0043, Japan}
\author{Mikito Koshino}
\affiliation{Department of Physics, Osaka University, Toyonaka, Osaka 560-0043, Japan}
\date{\today}

\begin{abstract}
We theoretically study emergent edge- and corner- localized states in monolayer black phosphorene.
Using the tight-binding model based on the density functional theory,
we find that the multi-orbital band structure due to the non-planar puckered geometry
 plays an essential role in the formation of the boundary localized modes.
In particular, we demonstrate that edge states emerge at a boundary along an arbitrary crystallographic direction,
and it can be understood from the fact that the Wannier orbitals associated with $3p_x,3p_y,3p_z$ orbitals occupy all the bond centers of phosphorene.
At a corner where two edges intersect, we show that multiple corner-localized states appear
due to hybridization of higher-order topological corner state and the edge states nearby.
These characteristic properties of the edge and corner states can be intuitively explained by a simple topologically-equivalent model
where all the bond angles are deformed to 90$^\circ$.
\end{abstract}

\maketitle
\section{introduction}
\label{sec_intro}

\begin{figure}
		\begin{center}
		\includegraphics[width=0.9\linewidth]{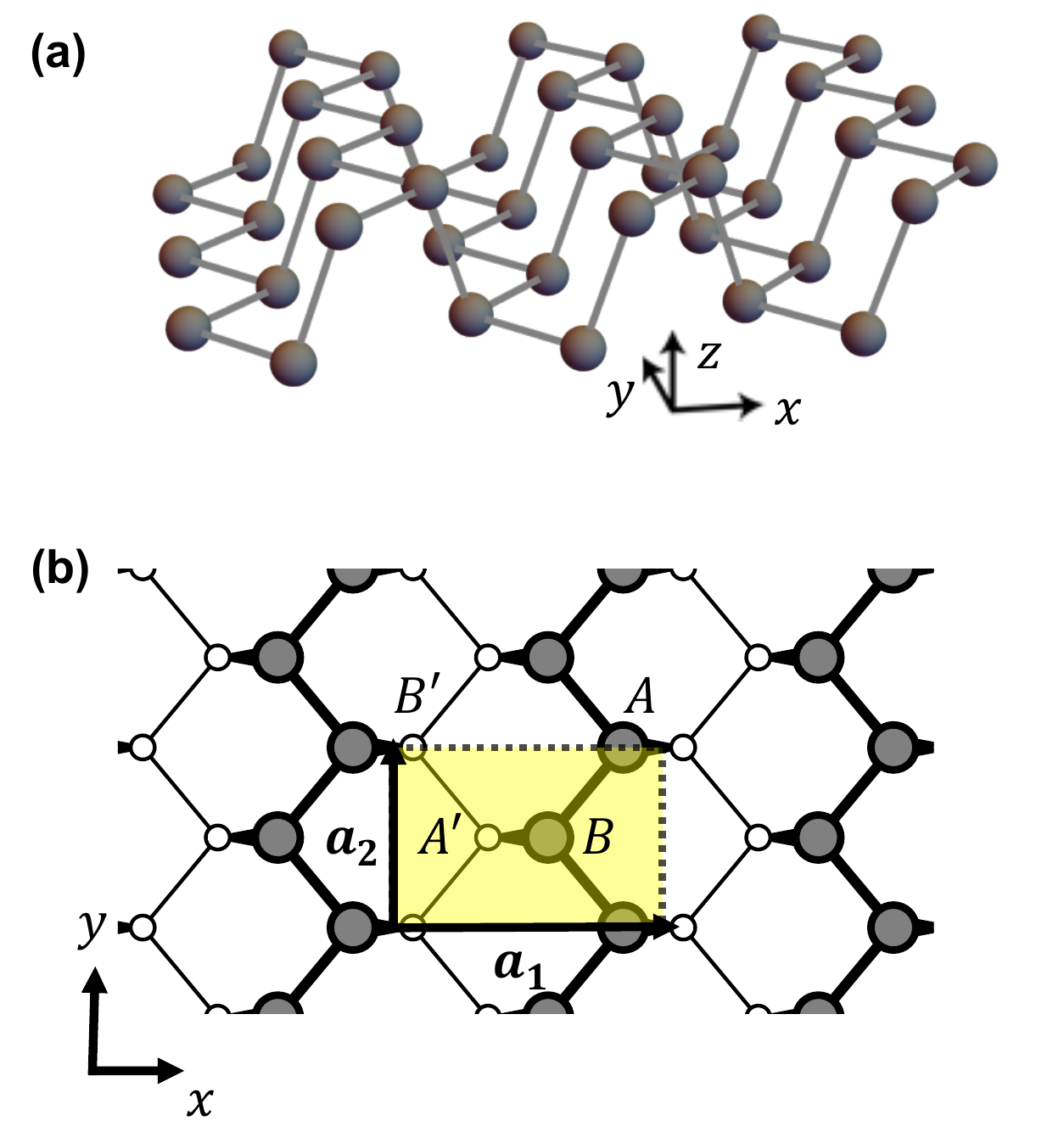}
		\caption{
		(a) Lattice structure of phosphorene.
		 (b) The top view. Gray (white) circles indicate atoms on the top (bottom) layers, 
		 and the yellow square is the unit cell with $A$, $B$, $A'$, and $B'$ sublattices.
		}
		\label{phos_atom}
		\end{center}
	\end{figure}

Black phosphorus is an allotrope of phosphorus with a van der Waals layered structure, which was discovered more than a century ago \cite{Bridgman1914}.
Its monolayer counterpart, black phosphorene  (hereafter just referred to as phosphorene)
~\cite{Liu2014,Li2014,Lu2014,Reich2014,Gomez2014,doi:10.1063/1.4868132,doi:10.1021/nl5032293,Peng2014,https://doi.org/10.1063/1.4885215,Dai2014,PhysRevB.89.235319,Qiao2014,Fei2014,Xia2014,doi:10.1021/acsami.5b10734, doi:10.1021/acsnano.5b01143,doi:10.1021/acsnano.5b02599,https://doi.org/10.1002/adfm.201502902,Kim2015,Akhtar2017,Watts2019},
have recently attracted considerable attention as a stable two-dimensional (2D) semiconductor
with potential applications for electronic devices \cite{doi:10.1021/nl5032293,doi:10.1063/1.4868132,Dai2014,Kim2015}.
Phosphorene has a puckered honeycomb structure shown in Fig.~\ref{phos_atom},
which is in contrast to graphene with a flat honeycomb lattice.
There the formation of $sp^3$ hybrids due to the non-planar structure
leads to a semiconducting band structure with an energy gap.

A notable difference between the phosphorene and graphene appears also in the edge states.
In graphene, edge states with a flat dispersion emerges at zigzag edges, while not at armchair 
edges~\cite{PhysRevB.54.17954,doi:10.1143/JPSJ.65.1920}.
For phosphorene, on the other hand, the previous theoretical works showed that the edge states emerge at both 
zigzag edges (along $y$ in Fig.~\ref{phos_atom}) and armchair edges (along $x$)~\cite{Carvalho2014,doi:10.1021/jp505257g,doi:10.1088/1367-2630/16/11/115004,Peng2014appl,Li2014jphyschem,doi:10.7566/JPSJ.84.013703,doi:10.7566/JPSJ.84.121004,PhysRevB.93.245413},
and the study was also extended to edges in diagonal directions~\cite{PhysRevB.93.245413}.
It was also pointed out that the edge states in phosphorene influence the electronic transport~\cite{Zhang2014,Amini2019} 
and performance as a electrocatalyst for the hydrogen evolution reaction~\cite{doi:10.1021/acs.chemmater.9b03031}.

Although the emergence of edge states in graphene was explained in terms of Zak phase~\cite{PhysRevLett.62.2747, delplace2011,PhysRevLett.89.077002},
those in phosphorene are not fully understood from a topological point of view.
For instance, the zigzag edge states of phosphorene was explained by the graphene-like minimal model only considering $p_z$ orbital~\cite{doi:10.1088/1367-2630/16/11/115004,PhysRevB.93.245413},
while the model does not explain the armchair edge states.
Interestingly, it was also shown that phosphorene supports a corner state at an intersection of different edges, within the same $p_z$
minimal model calculation \cite{PhysRevB.98.045125}. It was attributed to a higher-order topological property, which ensures the existence of  ($D-2$)-dimensional boundary-localized states in $D$-dimensional bulk system ~\cite{PhysRevB.98.045125,PhysRevB.77.235411,Wu_2010, Zhang2015, Song2017,PhysRevB.95.165443,PhysRevLett.119.246401,  PhysRevLett.120.026801, Serra-Garcia2018, Peterson2018, https://doi.org/10.1038/s41567-018-0224-7, PhysRevLett.123.186401, PhysRevLett.122.233902, PhysRevB.99.245151, PhysRevLett.123.216803,   PhysRevB.102.104109, Peterson2020, lee2020, takahashi2021}.

	\begin{figure*}[t]
		\begin{center}
		\includegraphics[width=1.\textwidth]{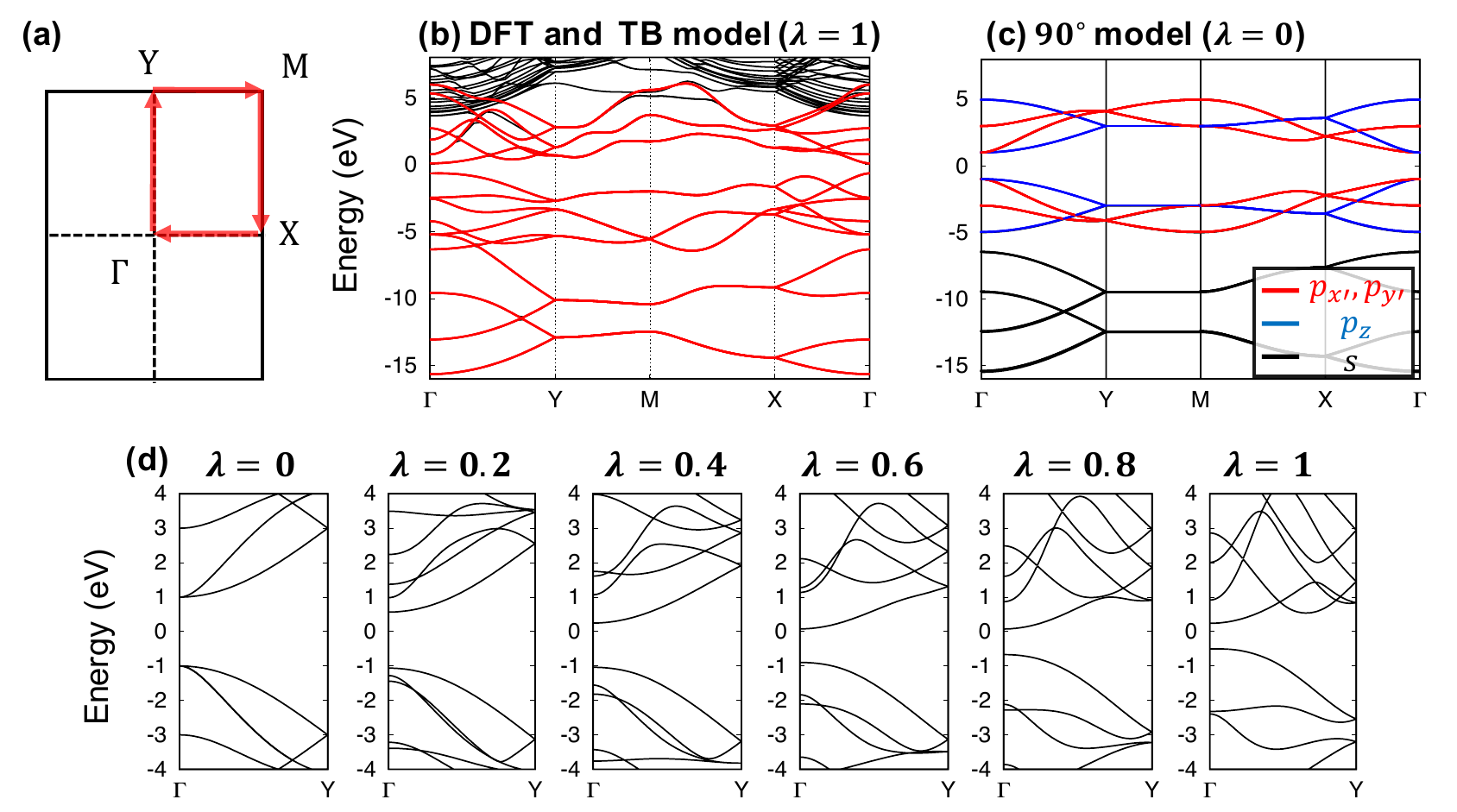}
		\caption{
		(a) The first Brillouin zone of phosphorene and the band-calculation path.
		(b) Band structure of the original DFT calculation (black curves) and that of the DFT-based tight-binding model (red). 
		The red curves almost completely overlap with the black curves in the low-energy region, leaving the free-electron bands in the high energy region.
		(c) Band structure of the $90^{\circ}$ model. 
		In (c), the red curves are the bands of $p_{x'}$ and $p_{y'}$ (degenerate on the path), 
		and the black and blue curves are of $s$ and $p_{z}$, respectively.
		(d) Deformation of  the band structures around Fermi energy from the $90^{\circ}$ model ($\lambda = 0$) to the DFT-based  tight-binding model ($\lambda = 1$). The band gap does not close.
		}
		\label{band_0_to_1}
		\end{center}
	\end{figure*}

On the other hand, the phosphorene is essentially a multi-orbital system, where the electronic bands around Fermi energy 
are composed of $3s, 3p_x, 3p_y, 3p_z$ orbitals hybridized by the puckered structure,
similarly to other non-planer 2D materials~\cite{Hattori2017}.
This is in sharp contrast to graphene where the low-energy states are dominated by $p_z$ orbital not participating in $sp^2$
bonding. Therefore, the edge / corner states of phosphorene and the associated topological nature should be considered in an
appropriate multi-orbital model.

In this paper, we study the edge and corner states of phosphorene using a multi-orbital tight-binding model based on
the density functional theory (DFT).
We find that the multi-orbital nature of phohsphorene forces the emergence of edge states in boundaries with {\it any} orientation:
The zigzag edge states are mainly contriubuted by $p_z$ orbital consistently with the $p_z$ minimal model \cite{doi:10.1088/1367-2630/16/11/115004,PhysRevB.93.245413},
while the armchair edge states turn out to be coming from $p_x$ and $p_y$ orbitals.
We also consider different types of edge terminations,
and observe that the multi-orbital nature gives extra edge states in addition to ones from $p_z$ orbital~\cite{PhysRevB.93.245413}.

The ubiquitous edge-state nature in phosphorene can be understood in terms of the center position of the Wannier orbital,
which is a topological invariant.
Here we show that, in phosphorene, the Wannier orbital reside in the middle of every single bond,
and hence cutting any bonds results in the half-breaking of the Wannier states and the emergence of the in-gap boundary states.
We also demonstrate that these characteristic properties can be analytically understood by using a topologically-equivalent $90^\circ$ model,
where all the bond angles are deformed to $90^\circ$.
There the edge states can be described as topological zero-energy modes.

At a corner where two edges intersect, we find that the higher-order-topological corner state is not stand-alone but 
inevitably hybridized with the edge states around the corner.
As a result of hybridization of the edge and corner orbitals, 
we have multiple corner-localized states composed of different orbitals.
These hybrid corner states can be explained using an edge-corner composite model only considering the unpaired edge and corner orbitals.
We conclude that the phosphorene is a unique material where the edge states and the corner states coexist and interact with each other.

The remaining sections are organized as follows. 
In Sec.~\ref{sec_bp}, we perform the DFT calculation and 
derive the multi-orbital tight-binding model of phosphorene. 
In addition, we introduce the $90^\circ$ model and its detailed properties.
In Sec.~\ref{sec_edge}, we reveal the existence of the edge states in various types of edge termination,
and correspondence to the fractionalization of the Wannier orbital. 
In Sec.~\ref{sec_corner}, we consider a finite-sized phosphorene flake and find the multiple corner states,
and clarify their origin by the edge-corner composite model. 
Finally, we give conclusions in Sec.~\ref{sec_con}.

\section{Model}
\label{sec_bp}

\subsection{DFT-based tight-binding model}
\label{sub_tb}
Phosphorene is a puckered honeycomb lattice of phosphorus atoms [Fig.~\ref{phos_atom}]. The primitive lattice vectors 
are given by $\bm{a}_1=(a,0,0)$ and  $\bm{a}_2=(0,b,0)$ with the lattice constants $a=4.476$ \AA\ and $b=3.314$ \AA. 
A unit cell consists of four phosphorus atoms labeled by $A$, $B$, $A'$, and $B'$, 
which are located at $\bm{\tau}_A=[(1/2-u)a,b/2,c]$, $\bm{\tau}_B=(ua,0,c)$, $\bm{\tau}_{A'}=-\tau_B$, and $\bm{\tau}_{B'}=-\bm{\tau}_A+\bm{a}_2$, respectively, with respect to the midpoint of the $A'B$ bond.  Here we defined $u=0.08056$ and $c=1.0654$ \AA~\cite{doi:10.1143/JPSJ.50.3362}.
The structure belongs to the non-symmorphic space group $Pmna$, which is generated by 
spatial inversion about the mid point of $A'B$,  a twofold rotation along $y$-axis, and the glide operation, i.e., the combination of half translation $\bm{t}=(\bm{a}_1+\bm{a}_2)/2$ and mirror reflection with respect to the $xy$ plane.

We calculate the electronic band structure by using \textit{ab initio} density functional theory (DFT) implemented in the QUANTUM ESPRESSO package. 
We employ the ultrasoft pseudopotentials with Perdew-Zunger selfinteraction corrected density functional, the cutoff energy of the plane-wave basis 30 Ry, and the convergence criterion of $10^{-10}$ Ry in $12\times12\times1$ $\bm{k}$-points mesh. 
The black curves in Fig.\ \ref{band_0_to_1}(b) show the resulting band structure along the high symmetry lines of the first Brilloiuin zone [Fig.1(a)],
where the Fermi energy is $E=0$. The valence band edge is located at the $\Gamma$ point where the band gap is $0.645$~eV.
All bands along $YM$ and $MX$ paths (Brillouin zone edge) are twofold degenerate (per spin) because of the glide symmetry. 

We derive a tight-binding model based on the DFT band structure. Since the energy bands near the Fermi energy
are dominated by the $3s$ and $3p$ electrons of phosphorus, we take into account the $s$, $p_x$, $p_y$, and $p_z$ orbitals at each of four atomic sites in the unit cell, giving 16 orbitals in total. 
Here we use the WANNIER90 package \cite{wannier90} 
and obtain the localized Wannier wavefunctions and the associated 
hopping parameters (see, Appendix~\ref{sec:hop} and Suplemental Material~\cite{sm}). 
The band structure of the derived tight-binding model is shown as red lines in Fig.~\ref{band_0_to_1}(b), which precisely reproduces
the original DFT energy bands.

\subsection{$90^{\circ}$ model}
\label{sub_90}
\begin{figure}
		\begin{center}
		\leavevmode
		\includegraphics[width=1.0 \hsize]{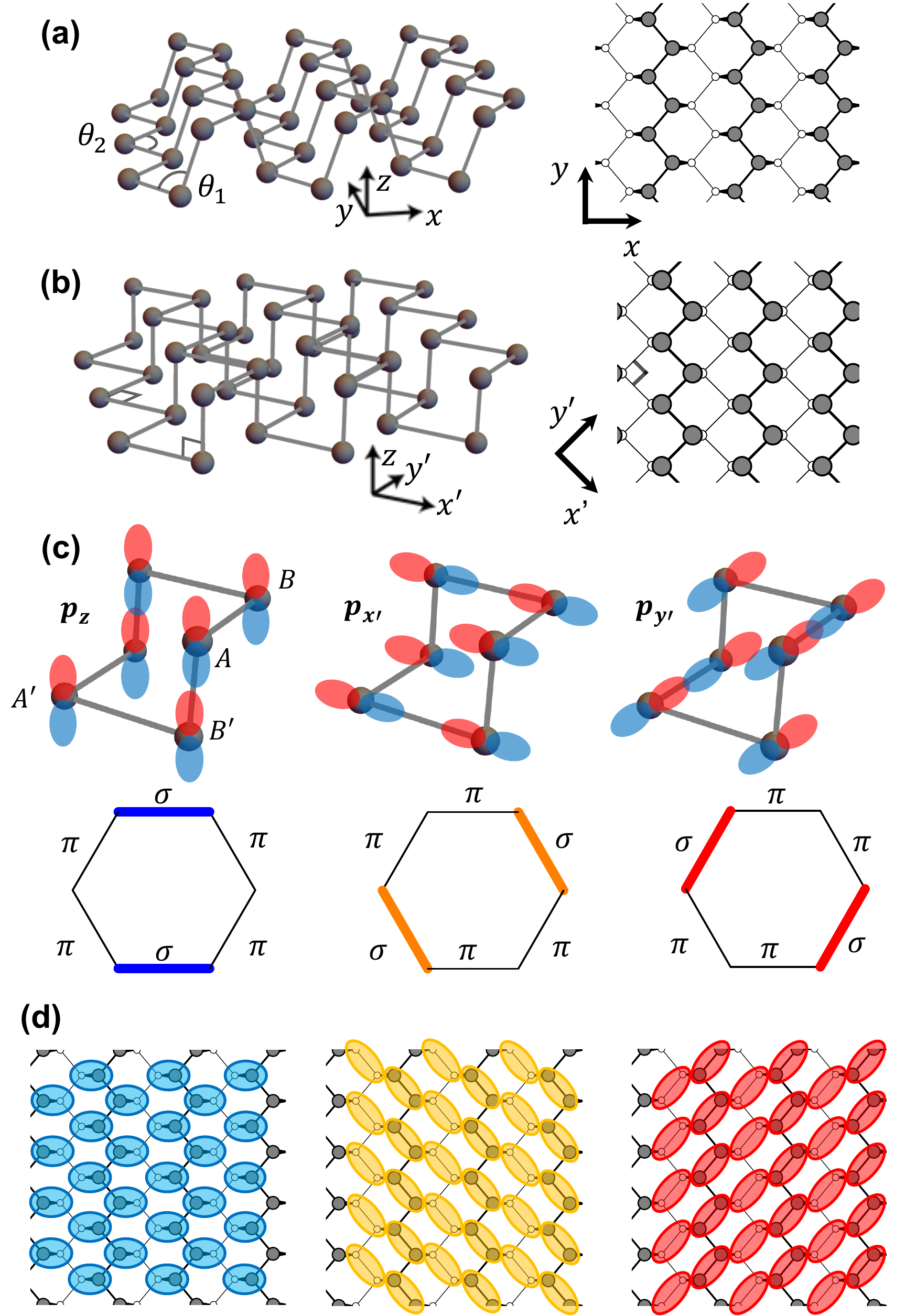}
		\caption{
		Atomic structure of (a) the original phosphorene and (b) $90^{\circ}$ model. The three-dimensional angle between bonds, $\theta_1\approx103^{\circ}$ and 
		$\theta_2\approx 98^{\circ}$ in (a) are deformed to $90^{\circ}$ in (b). (c) Orbital bases of $p_z$, $p_{x'}$, and $p_{y'}$ in the $90^{\circ}$ model, and the corresponding anisotropic honeycomb lattices where thick bonds indicate stronger hopping.
		(d)The Wanier orbitals associated with $p_z$, $p_{x'}$, and $p_{y'}$.
		}
		\label{90deg_basis}
		\end{center}
	\end{figure}

We introduce a simplified model which is topologically equivalent to the DFT-based tight-binding model for phosphorene.
The model is defined by deforming the angle between bonds  $\theta_1\approx103^\circ$ and $\theta_2\approx98^\circ$ 
[Fig.~\ref{90deg_basis}(a)] to  $90^{\circ}$ [Fig.~\ref{90deg_basis}(b)], so that
$A$ and $B’$ ($A’$ and $B$) are vertically aligned.
We also neglect all the further hoppings other than the nearest neighbor hoppings indicated by the bond lines in Fig.~\ref{90deg_basis}(b).
Note that all the crystalline symmetries inherent to the phosphorene, including glide, are preserved.

The resulting model (referred to as the $90^{\circ}$ model hereafter) is much simpler than the original model. 
Here we set the $x'$ and $y'$ axes parallel to the bond directions by rotating $x$ and $y$ axes by $45^\circ$,
and take $p_{x'}$ and $p_{y'}$ as the basis of in-plane $p$ orbitals [Fig.~\ref{90deg_basis}(c)].
In this basis, hopping between different $p$-orbitals $(p_{x'},p_{y'},p_z)$ becomes exactly zero, 
due to the orthogonality of the $p$ orbital orientations. 
We also neglect the hopping between $s$ and $p$ orbitals, because the energy bands originating from $s$-orbitals
are located far below in energy and the coupling hardly affect the states at the Fermi energy.

Then the Hamiltonian matrix ($16\times 16$) is written in a block diagonal form,
	\begin{eqnarray}
		H_{90^\circ}(\GVec{k})={\rm diag}[H_s(\GVec{k}),H_{x'}(\GVec{k}),H_{y'}(\GVec{k}),H_{z}(\GVec{k})],
		\label{H90}
	\end{eqnarray}
where the subscripts $s, x', y', z$ stand for $s, p_{x'}, p_{y'}, p_z$-orbitals, respectively.
The $4\times4$ block matrix $H_i(\GVec{k})\,(i=s, x', y',z)$ is written in the basis of $A$, $B$, $A'$, and $B'$ as
	\begin{eqnarray}
		H_i(\GVec{k}) =  
		\begin{pmatrix}
		0 & h_i(\GVec{k}) & 0 & t_{i,z} \\
		h^{\ast}_i(\GVec{k}) & 0 & t_{i,z} & 0 \\
		0 & t_{i,z} & 0  & h_i(\GVec{k})  \\
		t_{i,z} & 0 & h^{\ast}_i(\GVec{k})  & 0 \\
		 \end{pmatrix}
		 +\varepsilon_i,
		 \label{H_90_block}
	\end{eqnarray}
with
	\begin{eqnarray}
		h_i(\GVec{k}) &=&  t_{ix'} e^{{\rm i}\bm{k}\cdot\Delta\bm{r}_{x'}}+  t_{iy'} e^{{\rm i}\bm{k}\cdot\Delta\bm{r}_{y'}},\\
		t_{ij} &=&
		\begin{cases}
			t_s \quad (i=s)\\
			\delta_{ij}t_{\sigma}+(1-\delta_{ij})t_{\pi} \quad (i=x',y',z).
		\end{cases}
		 \label{h_90_ele}
	\end{eqnarray}
Here $t_{s} = 1.5$ eV is the hopping parameter for $s$-orbital,
$t_{\sigma} = 3$ and $t_{\pi} = 1$ are those for $\sigma$ and $\pi$ bonds of $p$-orbitals, respectively,
and $\Delta\GVec{r}_{x'}=a' \GVec{e}_{x'}$, $\Delta\GVec{r}_{y'}=a' \GVec{e}_{y'}$,
with $a'$ being the interatomic distance between $A$ and $B$ sites.
In the second term of Eq.\ (\ref{H_90_block}),
$\varepsilon_s = -11$~eV and $\varepsilon_p (=\varepsilon_{x'}=\varepsilon_{y'}=\varepsilon_{z}) = 0$ are the relative onsite potentials 
for $s$ and $p$ orbitals, respectively.
The band parameters $t_i, \varepsilon_i$ are determined so as to reproduce the approximate band structure of phosphorene.

The effective Hamiltonian (\ref{H_90_block}) is analytically solvable with the help of the glide symmetry 
\begin{eqnarray}\label{glidesym}
	G_{i}H_{i}(\bm{k})G_{i}^{\dag} = H_{i}(\bm{k}), 
\end{eqnarray}
with the glide operator
\begin{eqnarray}
	G_i=\eta_i
	\left(\begin{array}{cccc}
		0 & 0 & 1 & 0 \\
		0 & 0 & 0 & 1 \\
		1 & 0 & 0 & 0 \\
		0 & 1 & 0 & 0 
	\end{array}\right),
\end{eqnarray}
where $\eta_{s}=\eta_{x'}=\eta_{y'}=1$ and $\eta_{z}=-1$ stand for the mirror eigenvalues of the corresponding orbitals, 
with respect to the $x'y'$ plane. 
The symmetry Eq.~(\ref{glidesym}) decouples the Hamiltonian into two sectors with different eigenvalues of 
glide operator as $U_iH_i(\bm{k})U^\dag_i=\mathrm{diag}(H_{i,+}, H_{i,-})$, where $U_i$ is unitary matrix diagonalizing $G_i$. 
The Hamiltonian for each sector is given by
\begin{eqnarray}\label{glidesect}
	H_{i,\pm}(\bm{k})=
	\left(\begin{array}{cccc}
		0 & h_i(\bm{k})\pm \eta_i t_{i,z}  \\
		h_i^\ast(\bm{k}) \pm \eta_i t_{i,z} & 0  
	\end{array}\right) + \varepsilon_i, 
\end{eqnarray}
where $\pm$ is the eigenvalue of glide operator. 
The eigen energies are given by
\begin{eqnarray}\label{90eig}
	E_{i,\pm,s}= s |h_i(\bm{k}) \pm \eta_i t_{i,z}| + \varepsilon_i,
\end{eqnarray}
where $s=\pm$ are the electron and hole branches, respectively.
The band structure of Eq.~(\ref{90eig}) is shown in Fig.~\ref{band_0_to_1}(c),
where the red curves are the energy bands from $p_{x'}$ and $p_{y'}$ orbitals,
and the black and blue curves are from $s$ and $p_{z}$, respectively.
Here all the bands on the zone boundary $YMX$ are doubly degenerate due to the glide mirror symmetry just as in the original phosphorene.
In Fig.~\ref{band_0_to_1}(c), we see that the red lines $(p_{x'}, p_{y'})$ 
are degenerate also in $\Gamma X$ and $Y\Gamma$, but it is an artifact of the 90$^\circ$ model which lacks mixing of $p_{x'}$ and $p_{y'}$ orbitals.
It is understood by considering that $H_{x'}$ and $H_{y'}$ have the same eigen energies because $p_{x'}$ and $p_{y'}$ are interchanged by 
the twofold rotation around the $y$ axis and mirror reflection with respect to the $zx$ plane.

The 90$^\circ$ model [Fig.~\ref{band_0_to_1}(c)] and of phosphorene [Fig.~\ref{band_0_to_1}(b)] are topologically 
equivalent, in that the Hamiltonian can be continuously deformed from one to another
without closing the energy gap at the Fermi energy.
To demonstrate it, we define the deformation of Hamiltonian as,
\begin{eqnarray}
	H_{\lambda}(\GVec{k}) = (1-\lambda)H_{90^\circ}(\GVec{k}) + \lambda H_{{\rm black}}(\GVec{k}),
	\label{90deg2black}
\end{eqnarray}
where $H_{90^\circ}$ and $H_{\rm black}$ represent 90$^\circ$ model and the DFT-based model, respectively,  and $0\le \lambda\le 1$ is the tuning parameter.
Figure~\ref{band_0_to_1}(d) presents the band structure in changing $\lambda$,
where we see that energy gap does not close between $\lambda=0$ (90$^\circ$ model) and $\lambda=1$ (DFT-based model),
i.e., there are no topological phase transition.

A topological invariant which is relevant in the current problem is the center position of the Wannier orbitals,
which is rigorously fixed in a continuous deformation of the system \cite{PhysRevX.7.041069,Song2017,Bradlyn2017,Cano2021,xu2021}.
The Wannier orbital center of the 90$^\circ$ model can be easily identified by the following argument.
The decoupled $p_{x'}$, $p_{y'}$, and $p_z$ blocks in Eq.~(\ref{H_90_block})
can be mapped to anisotropic honeycomb lattices as shown in Fig.~\ref{90deg_basis}(c).
For instance, the $p_z$ orbital [the leftmost panel]
forms $\sigma$ bonds along the $z$ direction while $\pi$ bonds on  the $xy$ plane.
As the hopping integral is larger for $\sigma$ bonds than for $\pi$ bonds  ($t_{\sigma}>t_{\pi}$),
the system is mathematically equivalent to a single-orbital tight-binding model on a flat honeycomb lattice,
where the hopping is stronger in one direction ($t_{\sigma}$) than the other two directions ($t_{\pi}$), as indicated as thick and thin bonds
in the lower-left panel of Fig.~\ref{90deg_basis}(c).
Likewise, the $p_{x'}$ and $p_{y'}$ blocks can also be mapped to anisotropic honeycomb models,
where the strongest bonds appear in different directions.

As a whole, the strong bonds from three different $p$ orbitals cover all the three inequivalent bonds in the honeycomb lattice.
For an anisotropic honeycomb model, generally, the energy spectrum is gapped when $t_{\sigma}>2t_{\pi}$
(which is true in our case), and then the Wannier center of the valence and conduction bands are located at the 
center of the strongest bond \cite{PhysRevB.80.153412}.
In the 90$^\circ$ model as a whole, therefore, a Wannier orbital center is located at the midpoint of every single bond, and 
 different orbitals characters correspond to different bond directions as shown in Fig.~\ref{90deg_basis}(d).
Because the 90$^\circ$ model and phosphorene are topologically equivalent,
we conclude that the phosphorene also has the Wannier centers in the middle of all the bonds.
Alternatively, we can also uniquely identify the Wannier center positions from the 
irreducible representations at the symmetric points in the Brillouin zone,
leading to the same result (see, Appendix \ref{sec:sym}).

A class of materials where 
the Wannier orbital centers are not located at atomic sites (but at the bond center, for example) 
is referred to the obstructed atomic insulator (OAI)~\cite{Bradlyn2017, Cano2021, xu2021}. 
In OAI, edge-localized modes emerge when the system is terminated at a boundary cutting though the Wannier orbital center.
If the Wannier orbital localizes at the corner of the system, particularly,
the system has a zero-dimensional corner state, which is regarded as a higher-order topological 
state in a 2D system \cite{PhysRevLett.119.246401, Serra-Garcia2018, PhysRevLett.120.026801, PhysRevLett.123.216803, PhysRevLett.122.233902,PhysRevB.98.045125, PhysRevB.99.245151, Peterson2020, Peterson2018, lee2020, takahashi2021, PhysRevB.102.104109}.
According to the above discussion, the phosphorene is an OAI where the three orbital sectors have Wannier centers at all different bond centers. 
In the following sections, we will show that various edge and corner states emerge in the phosphorene
depending on the boundary configuration, 
where the multi-orbital nature plays a crucial role.

\section{edge states}
\label{sec_edge}
	\begin{figure}
		\begin{center}
		\leavevmode
		\includegraphics[width=1.0 \hsize]{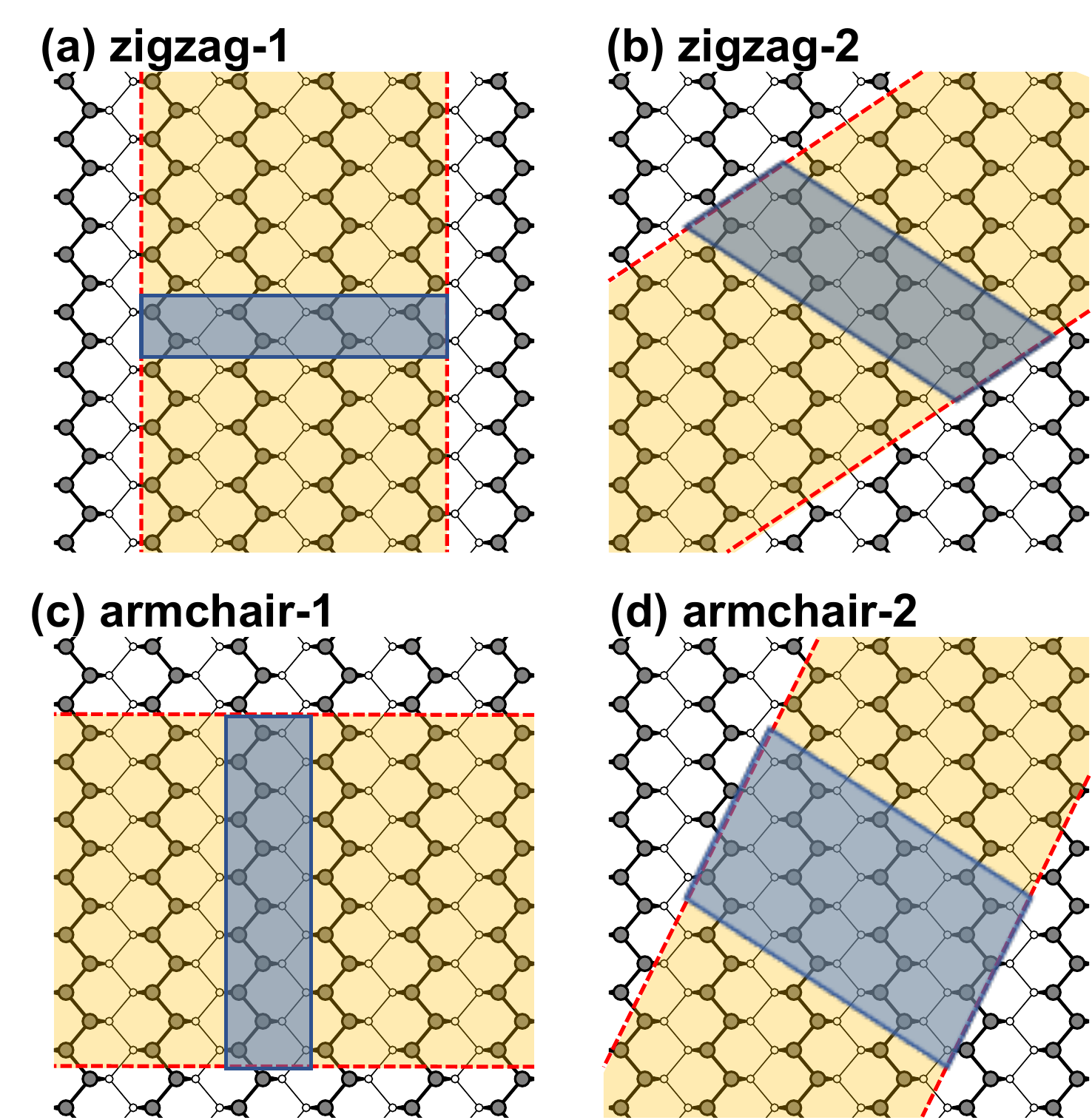}
		\caption{Four types of nanoribbons considered in this work. The yellow region represents the ribbon, and the blue parallelogram is the super unit cell of the ribbon.}
		\label{ribbon_atom}
		\end{center}
	\end{figure}

In this section, we calculate the energy band structures of phosphorene nanoribbons with various edge terminations.
We consider four different edge structures, 
zigzag-1, zigzag-2, armchair-1, armchair-2 illustrated in Figs.~\ref{ribbon_atom}(a) - \ref{ribbon_atom}(d), respectively,
which are parallel to $\bm{a}_2$, $\bm{a}_1+\bm{a}_2$, $\bm{a}_2$, and $\bm{a}_1+3\bm{a}_2$, respectively.
In the figure, the yellow region represents the ribbon, and the blue parallelogram is the super unit cell for the ribbon.
These four cases were previously studied in a minimal model only including $p_z$ orbital \cite{doi:10.1088/1367-2630/16/11/115004,PhysRevB.93.245413}, 
where the edge states are found only in the zigzag-1 and armchair-2 edges.
In the first principles study, on the other hand, the edge state was also found in the armchair-1 edges
\cite{doi:10.1021/jp505257g,Carvalho2014,doi:10.1088/1367-2630/16/11/115004,Peng2014appl,Li2014jphyschem,doi:10.7566/JPSJ.84.013703,doi:10.7566/JPSJ.84.121004}, while its origin is not well understood.
Below we systematically study all the types of edges, and relate the emergence of the edge states to the Wannier orbital position 
argued in the previous section.
We will see that the multi-orbital nature of the band structure gives rise to extra edge states missing in the $p_z$-only model.

	\begin{figure*}
		\begin{center}
		\leavevmode
		\includegraphics[width=1.0 \hsize]{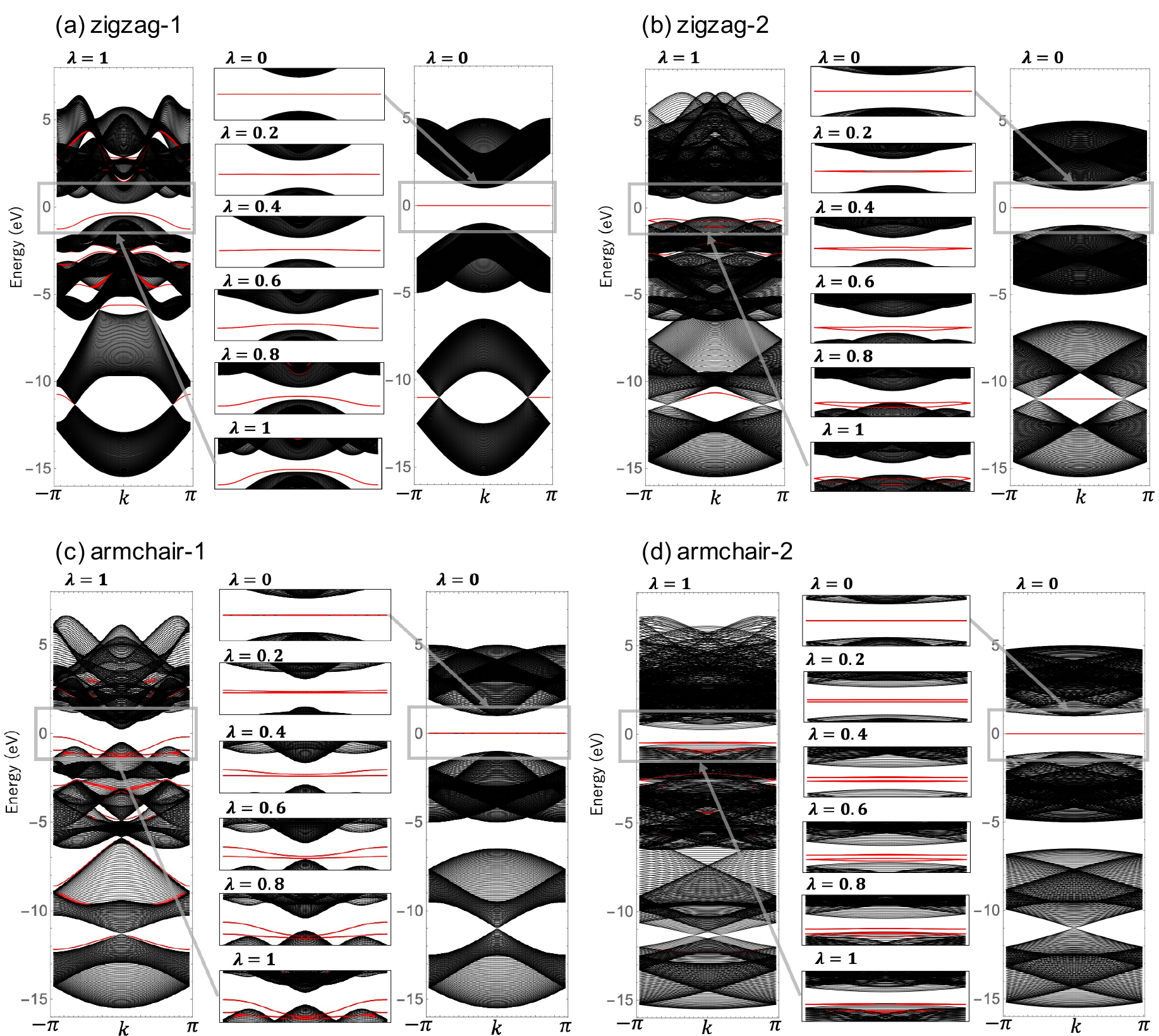}
		\caption{
		Band structures of four types of nanoribbons. In each figure, the left and right panels are the band structures of
 the DFT-based tight-binding model ($\lambda =1$) and the $90^{\circ}$ model ($\lambda =0$) , respectively. The middle row shows the evolution of the low-energy part in a continuous deformation from $\lambda =0$ to 1.
The edge states are connected to the chiral zero states in the $90^{\circ}$ model. 
}
		\label{ribbon_band}
		\end{center}
	\end{figure*}

We numerically calculate the band structures of the phosphorene nanoribbons 
using DFT-based tight-binding model introduced in Sec.~\ref{sec_bp}.
Note that the systems actually considered are much wider than the illustration in Fig.~\ref{ribbon_atom},
including 960 atomic sites per super unit cell for Figs.~\ref{ribbon_atom} (a) - \ref{ribbon_atom}(c) and 1920 sites for (d).
The band structures of the four types of nanoribbons are displayed in the left columns of Figs.~\ref{ribbon_band}(a)-(d).
The black and red curves represent bulk states and edge states, respectively.
Here the edge states are identified by the condition that 
more than 80\% of the total amplitude is concentrated within the width of  a single bulk unit cell from the edge.
We see that the edge states are not generally isolated from the bulk bands in energy but partially overlap with the bulk bands.

The edge states are more clearly distinguished in the 90$^\circ$ model introduced in Sec.\ \ref{sub_90}.
The right panels in Fig.~\ref{ribbon_band}(a)-(d) plot the energy bands of the $90^{\circ}$ model counterparts of the same nanoribbons,
and the middle panels show a continuous deformation from the 90$^\circ$ model ($\lambda=0$) to the original DFT-based model ($\lambda=1$).
In the 90$^\circ$ limit, we see that the edge state bands around the central gap all converge to $E= 0$.
This is caused by the chiral symmetry of the $90^{\circ}$ Hamiltonian,
\begin{eqnarray}\label{chiral}
	\rho_3 [H_{i}(\bm{k}) - \varepsilon_i] \rho_3 = -[H_{i}(\bm{k}) - \varepsilon_i], 
\end{eqnarray}
where $\rho_3 = {\rm diag}(1,-1,1,-1)$.
Under the chiral symmetry, the energy spectrum for $i=s$, $x'$, $y'$, and $z$ sectors 
are symmetric with respect to $E= \varepsilon_i$ as observed in the right columns of Fig.~\ref{ribbon_band}(a)-(d). 
The edge states at  zero energy are the chiral zero modes (eigenstates of $\rho_3$) \cite{PhysRevLett.89.077002} of the $p$-orbital sectors,
and therefore they are fixed to $\varepsilon_p(=0)$, and necessarily isolated from the bulk bands.
The number of edge state bands, $N_{\rm e}$, does not change during the deformation $0\le\lambda\le1$,
and it can be easily obtained by counting the number of zero energy levels at the $90^{\circ}$ model.
We find $N_{\mathrm e}=1, 2, 2, 4$ for zigzag-1, zigzag-2, armchair-1, and armchair-2 edges, respectively.

\begin{figure}
		\begin{center}
		\leavevmode
		\includegraphics[width=0.5 \hsize]{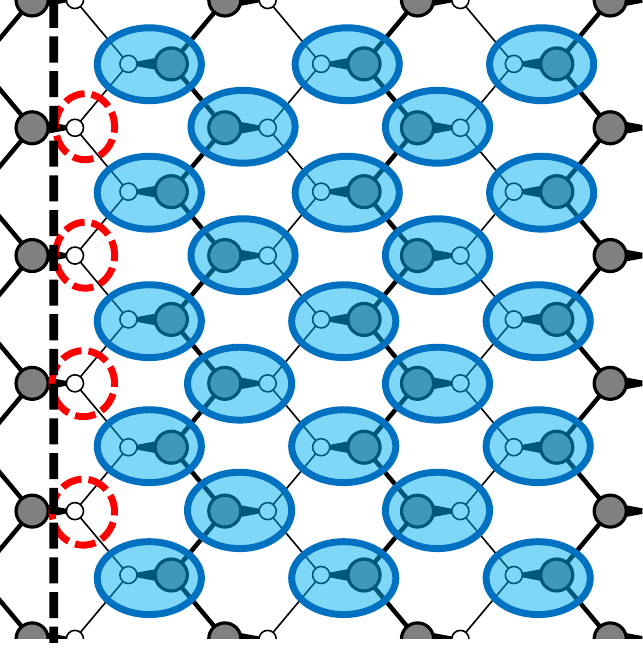}
		\caption{
		The Wannier states associated with $p_z$ in zigzag-1 termination (vertical dashed line). 
		Paired (Unpaired) orbitals are indicated by blue ellipse (red dashed circles).
		}
		\label{pz_cut}
		\end{center}
	\end{figure}

The emergence of the edge states can be intuitively understood from the fractionalization of the localized Wannier orbital,
in an analogous manner to the Su-Schrieffer-Heeger (SSH) model in one-dimension \cite{PhysRevLett.42.1698}.
As discussed in Sec.~\ref{sub_90}, every single bond in the phosphorene lattice is associated with the center of a single Wannier orbital.
In the zigzag-1 edge, for instance,
the edge line cuts the Wannier orbitals of the $p_z$ sector as illustrated in Fig.~\ref{pz_cut}.
Then the remaining uncoupled orbitals, indicated by dashed circles in Fig.~\ref{pz_cut},
form edge-localized states in the energy region outside the bulk bands.
This is actually the origin of the edge states of the zigzag-1,
and the number of the uncoupled orbital per super unit cell coincides with the number of the edge states, $N_{\rm e}=1$.
The Wannier orbital located at an interatomic bond corresponds to a covalent bond in chemistry,
and its broken half is nothing but a dangling bond.

\begin{figure}
		\begin{center}
		\leavevmode
		\includegraphics[width=1.0 \hsize]{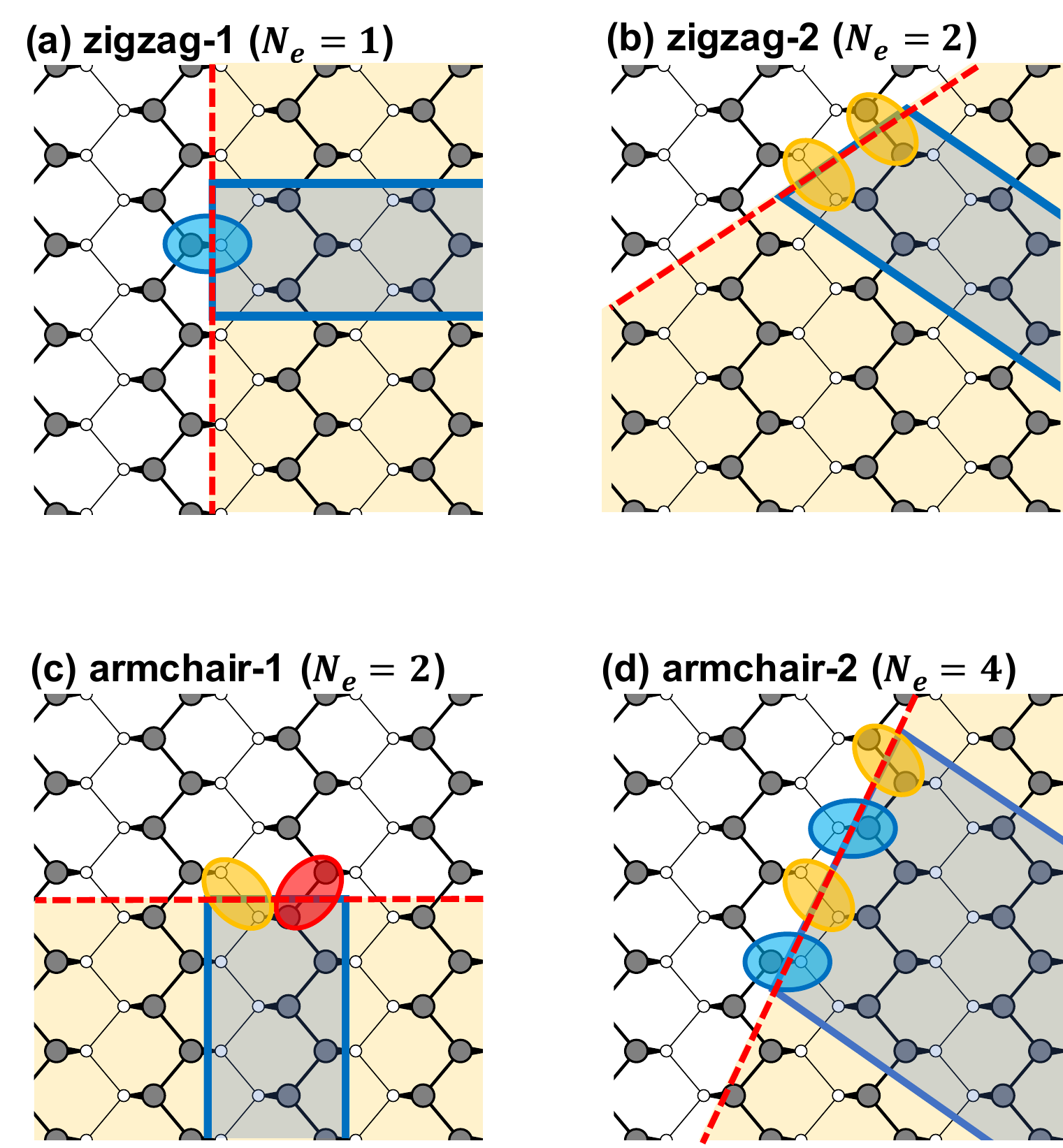}
		\caption{
Schematic illustration of the broken Wannier state for the four types of edge termination. 
Orange, red and blue ovals represent the broken Wannier states of $p_{x'}$, $p_{y'}$ and $p_z$ sectors, respectively.
The number of edge levels in the band calculation, $N_{\rm e}$, coincides with
the number of the broken Wannier states per a unit period of the ribbon.
}
		\label{edge_cut}
		\end{center}
	\end{figure}

The same analysis is applicable to the other edge structures as summarized in Fig.~\ref{edge_cut}. 
For the zigzag-2 nanoribbon [Fig.~\ref{edge_cut}(b)], 
we expect the two edge states since the edge line cuts the two Wannier functions of $p_{x'}$ orbital per super unit cell. 
In the same manner, the armchair-1 [Fig.~\ref{edge_cut}(c)] yields two edge states from $p_{x'}$ and $p_{y'}$ orbital,
and the armchair-2 [Fig.~\ref{edge_cut}(d)] yields four edge states from $p_{x'}$ and $p_{z}$ orbitals.
The results are all consistent with the number of edge state bands $N_{\mathrm e}$ in Fig.~\ref{ribbon_band}(a)-(d),
and also with the actual orbital character of the edge state wavefunctions.


In systems with time-reversal and space-inversion symmetries, 
the edge states originating from half-broken Wannier orbitals
can also be characterized by non-trivial Zak phase, which represents the charge polarization in the unit cell
~\cite{PhysRevLett.89.077002,Liu2017,Hughes2011,PhysRevB.84.195103,PhysRevB.99.245151,PhysRevB.88.245126,PhysRevB.95.035421,PhysRevB.96.235130,PhysRevB.101.161106,PhysRevResearch.2.033224}. 
Note that, however, the Zak phase only gives the parity of the number of edge states as it is $\mathbb{Z}_2$ valued,
so that the zigzag-2, armchair-1 and armchair-2 cases in our problem are all classified to  $\mathbb{Z}_2$-trivial.
On the other hand, the complete information of the Wannier orbital center unambiguously specifies 
the number of the edge states as well as their orbital characters as shown above.

It is also worth noting that, in Fig.\ \ref{ribbon_band}, the edge-state bands in the zigzag-2 and armchair-2 cases
always stick together at the Brillouin zone boundary ($k=\pm\pi$).
 This property is characteristic to the M\"obius twisted edge states protected by $\mathbb{Z}_2$ invariant,
which emerge in an edge-terminated system with glide symmetry~\cite{PhysRevB.91.155120,PhysRevB.93.195413}.
Note that the zigzag-2 and armchair-2 ribbons are glide symmetric because the edge direction
is parallel to the half-integer translation in the bulk glide operation. On the other hand, the glide symmetry is lost in the zigzag-1 and armchair-1 
termination, and the band sticking is absent accordingly.

In the zigzag-1 and zigzag-2 ribbons [Fig.~\ref{ribbon_band}(a) and (b)],
we also notice some edge states in  $E < -10$ eV far below the Fermi energy,
which are originating from $s$-orbital bands.
Similarly to the $p_z$ orbital of the graphene, 
the $s$-orbitals in the puckered honeycomb lattice form the Dirac cones as seen in Fig.~\ref{band_0_to_1}(b) (at $E\sim-11.48$ eV  on the $\Gamma Y$ line) 
and the zigzag edge states emerge between the two Dirac points just like in graphene \cite{doi:10.1143/JPSJ.65.1920}.

We expect that our tight-binding model captures the qualitative features such as the existence of the edge modes and its orbital character, which are the main scope of the paper. We note that the band dispersion of the edge-state band of zigzag-1 [Fig.\ \ref{ribbon_band}(a)] in our model 
qualitatively agrees with the results of full DFT calculations~\cite{Carvalho2014,Peng2014appl,doi:10.1021/jp505257g}.
In particular, if we calculate the energy bands of a narrower ribbon ($ \lesssim 30$ nm), we reproduce a small split of the edge-state bands at $k=0$ 
caused by the coupling of the two edges, which is also present in the DFT calculation.
For the armchair-1 ribbon [Fig.\ \ref{ribbon_band}(b)], our model is good enough to see the existence of the edge states, 
while we see some difference in the detailed band structure from the DFT results. We presume that the difference originates from a significant 
lattice relaxation at the armchair edge, which flattens the edge atoms in the puckered structure ~\cite{Carvalho2014,Peng2014appl,doi:10.1021/jp505257g}.
We leave further discussion on these effects to future study.

\section{corner states}
\label{sec_corner}


Now we consider a finite-sized phosphorene nanoflake to study the emergent corner states.
We employ the same strategy as for the edge states in the previous sections.
We calculate the energy spectrum and the eigenstates of a nanoflake
under the continuous deformation from  the 90$^\circ$ model ($\lambda=0$) to the DFT-based model ($\lambda=1$),
and clarify the physical origin of the obtained corner states.

	\begin{figure}
			\begin{center}
		\leavevmode
		\includegraphics[width=1.0 \hsize]{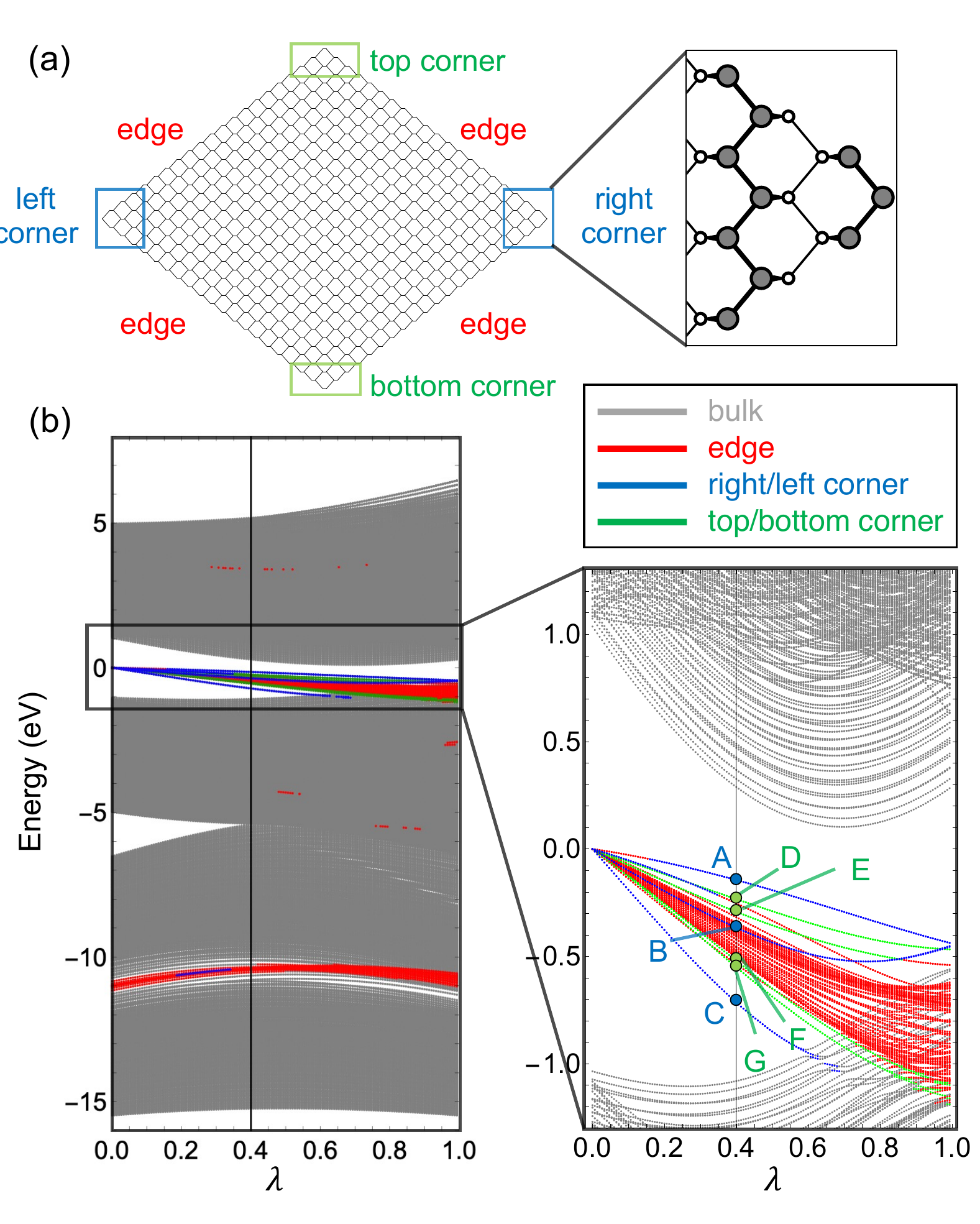}
		\caption{
(a) Atomic structure of a phosphorene flake considered in this work. 
(b) The evolution of the energy spectrum from the 90$^\circ$ model ($\lambda=0$) to the DFT-based tight-binding model ($\lambda=1$)
with the right panel showing the enlarged plot around the zero energy.
Blue (green) points indicate the left/right (top/bottom) corner states,
red points are the edge states, and gray dots are the bulk states. 
		}
		\label{corner_level}
		\end{center}
	\end{figure}

To be specific, we consider a phosphorene nanoflake as shown in Fig.~\ref{corner_level}(a), which includes 880 atoms.
Each corner is regarded as an intersection of two zigzag-2 edges,
and the left/right corners and the top/bottom corners have inequivalent structures with different intersecting angles.
Figure \ref{corner_level}(b) plots the energy spectrum of the flake as a function of $\lambda$,
where the right panel is the enlarged plot around the zero energy.
Here the blue (green) points indicate the left/right (top/bottom) corner states,
which have more than 60\% of the total amplitude within three unit cells from the corner site.
The red dots represent the edge states, which have more than 60\% amplitude
within a unit cell from the boundary (and which are not corner states).
The rest gray dots are the bulk states.
For the 90$^\circ$ model ($\lambda=0$),  all the edge and corner states are degenerate at $E=0$ 
because of the chiral symmetry, Eq.\ (\ref{chiral}).
With increasing $\lambda$, the degeneracy is lifted by breaking the chiral symmetry, 
and the corner states and edge states are resolved.


We first focus on the left/right corner states (blue dots) at $\lambda = 0.4$.
As seen in Fig.\ \ref{corner_level}(b), there are three branches of left/right corner levels, which are labeled as A, B, and C. 
In increasing $\lambda$, the level C eventually hybridizes with the bulk states, while the level A and B survive up to $\lambda=1$.
Here we consider $\lambda = 0.4$, because the origin of the corner states can be argued
without suffering from the complexity due to the hybridization with the bulk states.

\begin{figure}
		\begin{center}
		\leavevmode
		\includegraphics[width=1.0 \hsize]{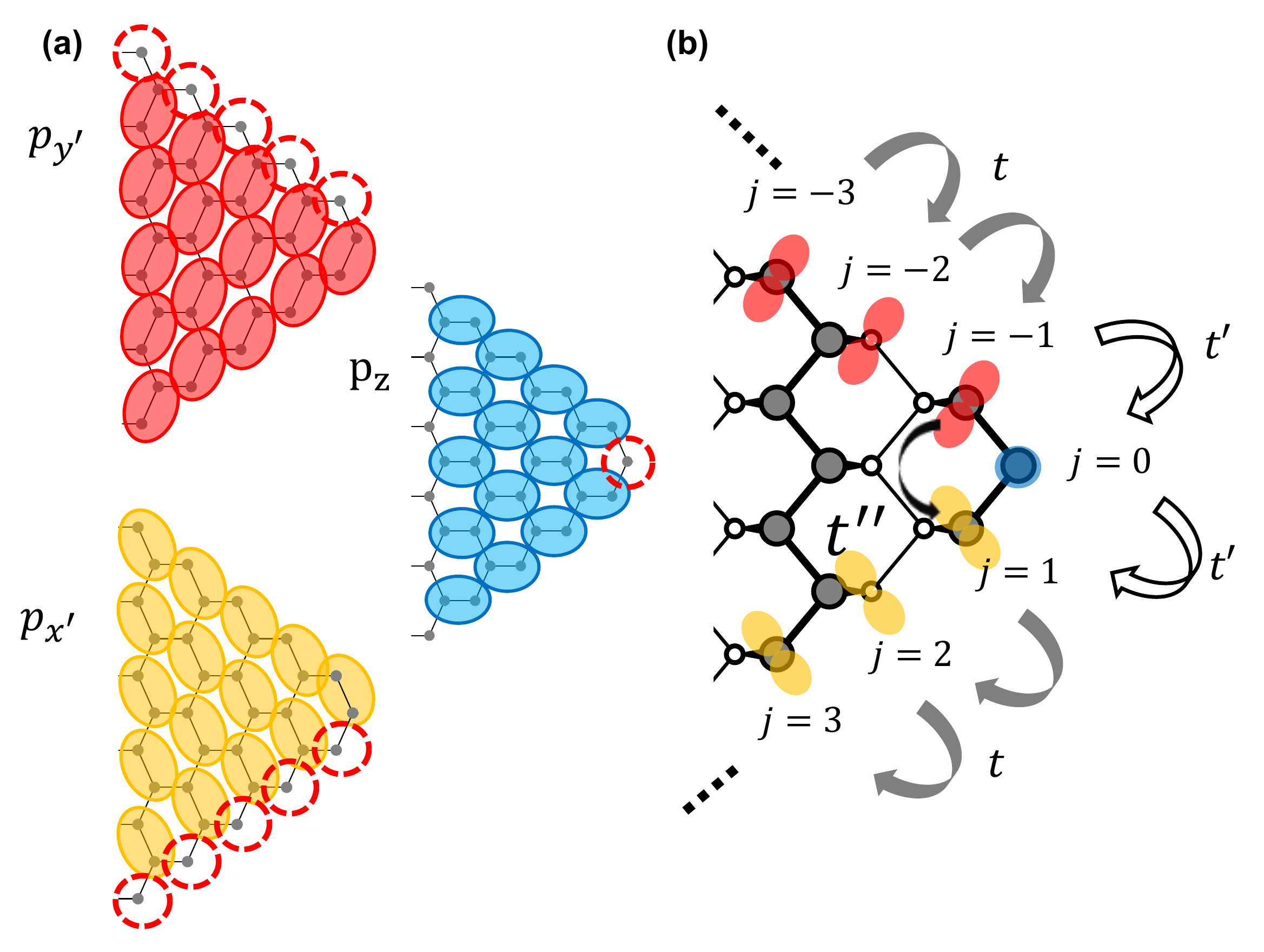}
		\caption{
		(a) Schematic illustration of Wannier orbital originating from the $p_x$, $p_y$, and $p_z$ orbitals 
		near the right corner of the phosphorene flake. The unpaired orbitals are marked by the red dashed circles.
		(b) Edge-corner composite model near the right corner, which is composed of the unpaired orbitals in (a). 
		}
		\label{corner_ec}
		\end{center}
	\end{figure}

\begin{figure*}
		\begin{center}
		\leavevmode
		\includegraphics[width=160mm]{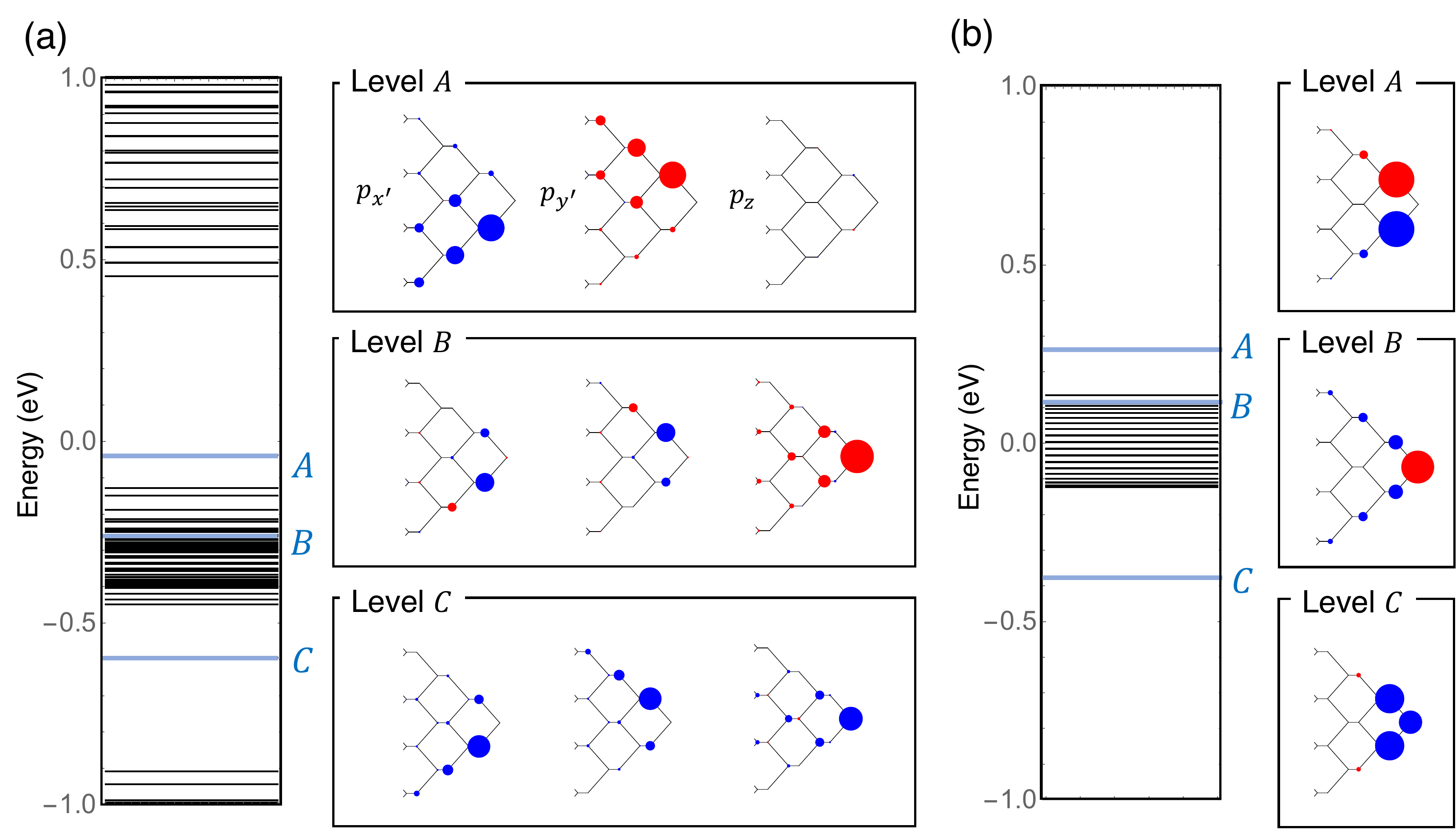}
		\caption{
		Energy levels and wavefunctions of the right-corner states $A$, $B$, and $C$ [Fig.\ \ref{corner_level}(b)] in
		the phosphorene flake with $\lambda=0.4$, obtained from (a) the original tight-binding model $H_{\lambda}^{\mathrm{flake}}$ [Eq.~\ref{H_nf}] 
		and (b) the corresponding edge-corner composite model $H_\mathrm{EC}$ [Eq.~\ref{ecmodel}]. 
		The radius and color of circles (red/blue) in the wavefunction indicate the amplitudes and phase (red / blue for plus / minus) of the wavefunction.
		}
		\label{corner_wf}
		\end{center}
	\end{figure*}

The presence of multiple corner states can only be explained by the multi-orbital picture.
Figure \ref{corner_ec}(a) illustrates the schematic picture for the Wannier orbitals of $p_{x'}$, $p_{y'}$ and $p_z$ sectors in the $90^\circ$ model
($\lambda =0$).
The dashed circles represent uncoupled orbitals which give the zero energy levels.
Here we notice that the $p_z$ sector has only a single uncoupled orbital at the corner,
while the $p_{x'}$ and $p_{y'}$ sectors have ones at all the outermost sites (hereafter referred to as edge sites) along the two edge lines.
In the minimal model only considering $p_z$,
a single corner orbital leads to a single corner state \cite{PhysRevB.98.045125}.
It is regarded as a higher-order topological insulator in that it has obstructed corner orbitals while not edge orbitals \cite{PhysRevB.98.045125,PhysRevB.77.235411,Wu_2010,PhysRevB.95.165443,PhysRevLett.119.246401,  PhysRevLett.120.026801, Serra-Garcia2018, Peterson2018, https://doi.org/10.1038/s41567-018-0224-7, PhysRevLett.123.186401, PhysRevLett.122.233902, PhysRevB.99.245151, PhysRevLett.123.216803,   PhysRevB.102.104109, Peterson2020, lee2020, takahashi2021}.
In the present multi-orbital phosphorene model, on the other hand, the $p_z$ corner state is hybridized with the edge zero modes of $p_{x'}$ and $p_{y'}$
when $\lambda$ is switched on, resulting in three corner states in total. 
Figure \ref{corner_wf}(a) shows the wavefunctions of these corner states at $\lambda=0.4$,
where we actually see that the wave amplitudes are mainly concentrated on those uncoupled orbitals.

To qualitatively understand the emergence of the multiple corner states, we introduce an effective edge-corner composite model
which takes account of only the uncoupled orbitals.
We label the edge and corner sites  by $j= 0,\pm1,\pm2 \cdots$ as in Fig.~\ref{corner_ec}(b),
where $j=0$ represents the $p_z$ orbital at the corner site,
and the positive (negative) $j$'s correspond to $p_{x'}$ ($p_{y'}$) orbitals at the lower (upper) edge.
We consider a one dimensional tight-binding model of these boundary orbitals to describe the in-gap states.
The Hamiltonian is explicitly written as 
\begin{align}
\label{ecmodel}
H_{\mathrm{EC}} = 
\Biggl(\sum^\infty_{j=-\infty} -t_{j+1,j} c^\dagger_{j+1} c_j + {\rm H.c} \Biggr) \,\,
-t'' (c^\dagger_{1} c_{-1}  + {\rm H.c})
\end{align}
with 
\begin{eqnarray*}
t_{j+1,j} = 
\left\{
\begin{array}{l}
t  \quad  (j\leq -2, j\geq 1),\\
t' \quad  (j=-1, 0),
\end{array}
\right.
\end{eqnarray*}
where $c^\dagger_j$ and $c_j$ are the electron creation and annihilation operators at site $j$, respectively,
$t$ is the hopping between the neighboring edge sites,
$t'$ is that between the edge site and the corner site,
and $t''$ is the second-nearest neighbor hopping between the edge site $j=\pm1$.
Here the corner site $j=0$ works as an impurity in a one-dimensional tight-binding chain of the edge sites.

The three hopping parameters $t$, $t'$, and  $t''$ in $H_{\rm EC}$ is determined 
by the second-order perturbation theory as follows.
First, we write the Hamiltonian of the phosphorene nanoflake (parameterized by $\lambda$) in a block form,
	\begin{eqnarray}
		H_{\lambda}^{\rm flake}=
		\begin{pmatrix}
		H_{0}  &  U \\
		U^{\dagger} &  H_{\rm bulk}
		 \end{pmatrix}
		\label{H_nf}
	\end{eqnarray}
where $H_{\rm 0}$ is the Hamiltonian projected on the edge and corner orbitals, $H_{\rm bulk}$ is that on the remaining orbitals, and 
$U$ is the coupling between them. 
In the projection to $H_0$, we take $p_{x'}$ and $p_{y'}$ on the edge sites to be parallel to the in-plane bonds of the real phosphorene, 
where the relative angle of $p_{x'}$ and $p_{y'}$ is $\theta_2=98^{\circ}$.
By treating $U$ as a perturbation, the effective Hamiltonian for the edge and corner orbitals is obtained by
	\begin{eqnarray}
		H_{\rm eff}= H_{\rm 0} + U^{\dagger} \frac{1}{E-H_{\rm bulk}} U, 
		\label{H_eff}
	\end{eqnarray}
where we take $E$ to be the average of eigenvalues of $H_{\rm 0}$.
Finally, $t$, $t'$, and  $t''$ can be extracted from the corresponding matrix elements of $H_{\rm eff}$. 
For example, the parameters for $\lambda=0.4$ are $t=0.060$ eV, $t'=0.15$ eV, and $t''=0.24$ eV.

\begin{figure}[b]
		\begin{center}
		\leavevmode
		\includegraphics[width=65mm]{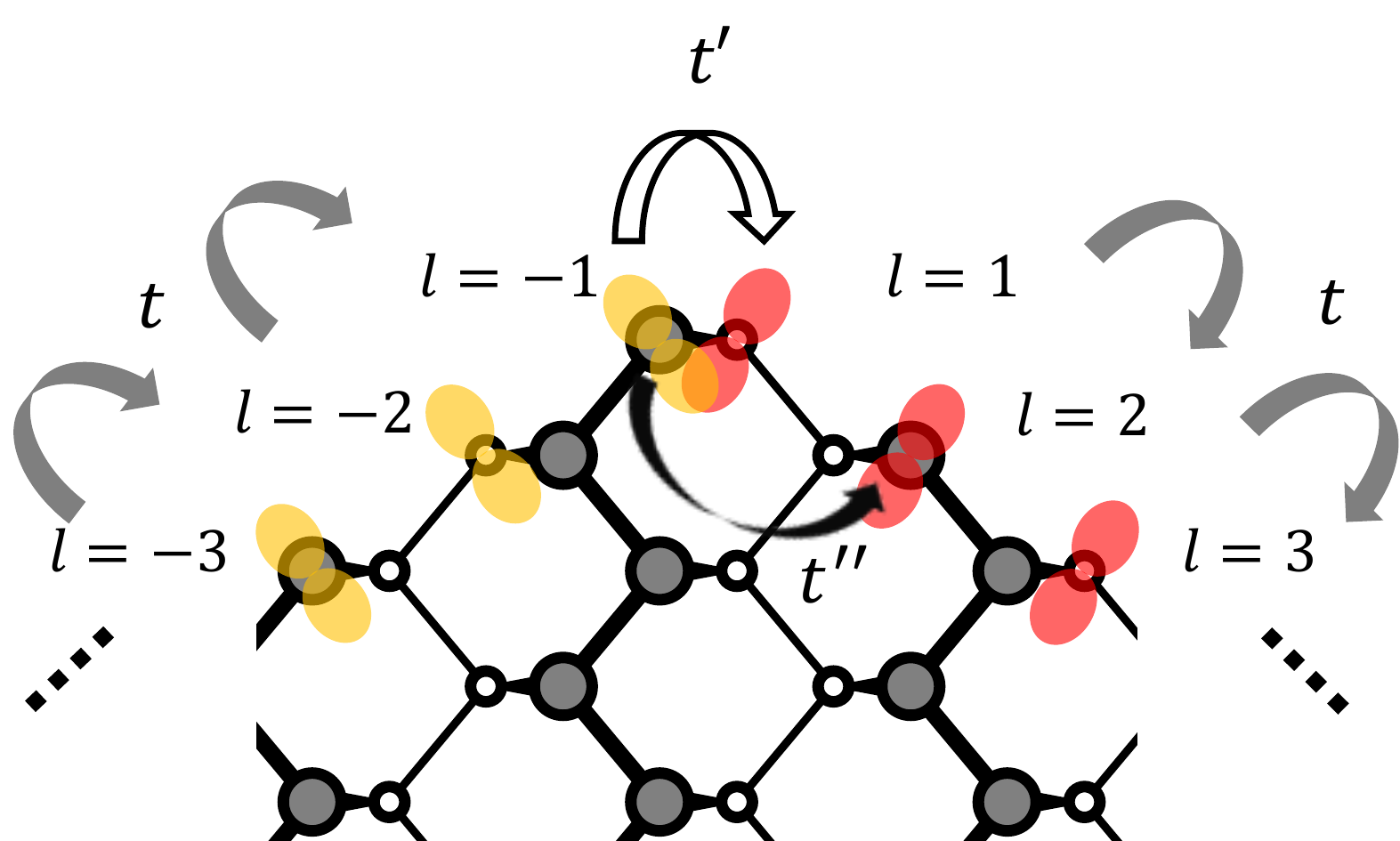}
		\caption{
		Edge-corner composite model near the top corner of the flake. 
		Uncoupled $p_{x'}$, $p_{y'}$ orbitals at the boundary are shown in yellow, and red respectively, and are labeled by $l$.
		}
		\label{corner_ec2}
		\end{center}
	\end{figure}

	\begin{figure*}
		\begin{center}
		\leavevmode
		\includegraphics[width=165mm]{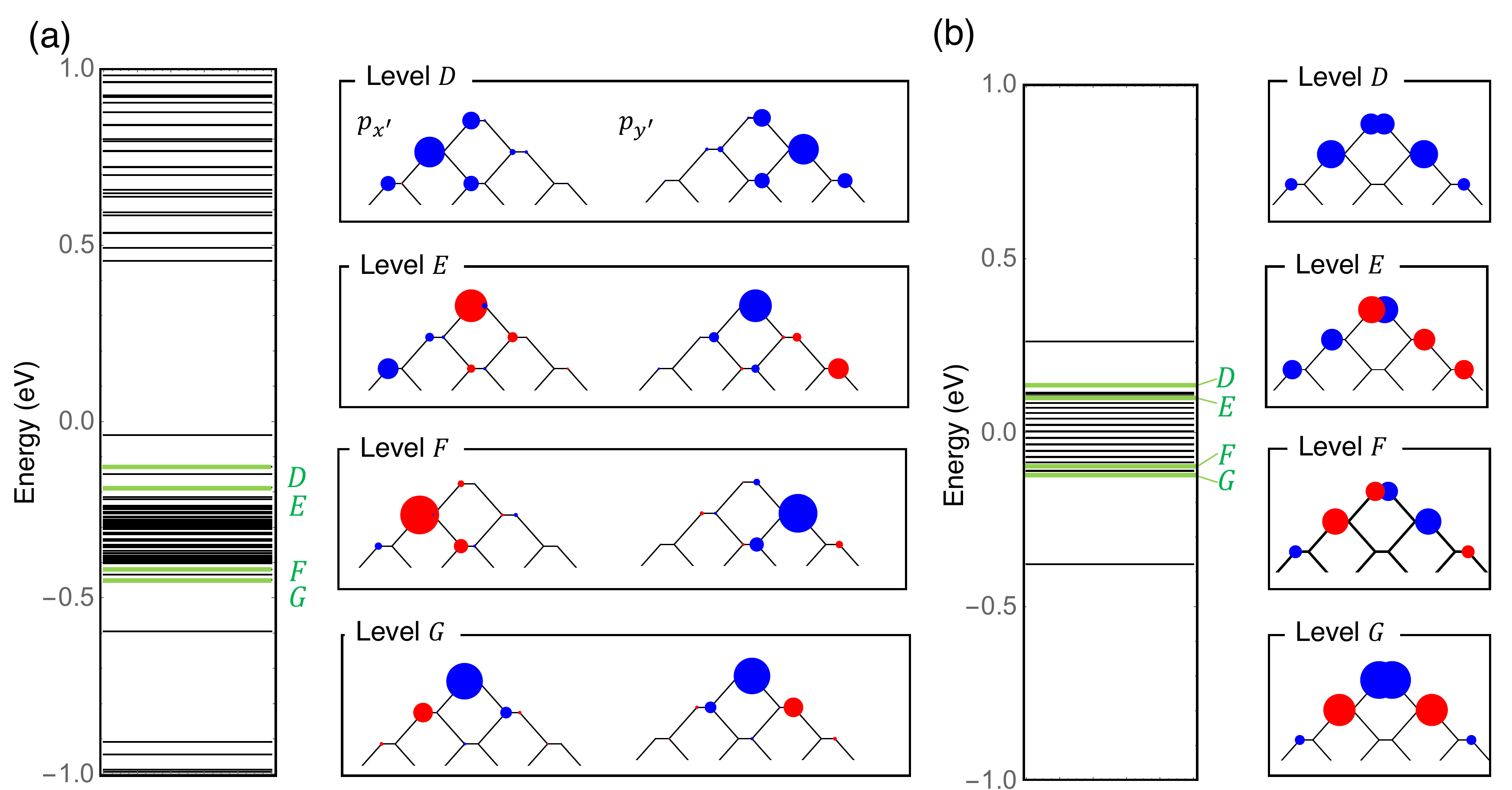}
		\caption{
		Plots similar to Fig.~\ref{corner_wf} for the top-corner states $D$, $E$, $F$ and $G$ [Fig.\ \ref{corner_level}(b)].
		}
		\label{corner_tb}
		\end{center}
	\end{figure*}

Figure \ref{corner_wf}(b) presents
the energy spectrum and the wavefunctions of the corner states
in the edge-corner composite model at $\lambda=0.4$.
In the calculation, we assumed a closed ring geometry
by connecting the upper and lower edge sites in a far away point, where the number of the total sites is 82.
We see that this simple model qualitatively reproduces the energy spectrum and wave function of the DFT-based model in Fig.\ \ref{corner_wf}(a).
The three eigenstates of $H_{\rm EC}$ marked as $A$, $B$, and $C$ are localized near the corner,
and the wavefunctions and their spatial symmetry agree with the corresponding states of the original model.
Therefore, the multiple corner states of phosphorene can be understood as a result of hybridization of the uncoupled edge and corner orbitals,
where the multi-orbital property is essential.



The interpretation using the effective edge-corner model is applicable also to the top/bottom corner states.
In the energy spectrum of Fig.\ \ref{corner_level}(b), there are four branches of top/bottom corner states (green dots) at $\lambda=0.4$,
which we label $D$, $E$, $F$, and $G$ in descending order of energy.
Figure \ref{corner_ec2} illustrates the uncoupled orbitals around the top corner.
Now we have $p_{x'}$ and $p_{y'}$ orbitals along the two edges,
while the corner-isolated orbital, like $p_z$ for the right corner, is absent in this case.
We label the $p_{x'}$ orbitals as $l = -1,-2,-3,\cdots$
and the $p_{y'}$ orbitals as $l = 1,2,3,\cdots$, where the site $l=0$ is missing.
The effective edge-site model is written as,
\begin{align}
\label{ecmodel2}
	H'_{\mathrm{EC}} = & 
	\Biggl[\sum^\infty_{l=1} -t(c^\dagger_{l+1} c_l + c^\dagger_{-(l+1)} c_{-l}) + {\rm H.c} \Biggr] \,\, \nonumber\\ 
&	- t' (c^\dagger_{1} c_{-1}  + {\rm H.c.}) - t'' (c^\dagger_{1} c_{-2}  + c^\dagger_{-1} c_{2} + {\rm H.c})							
\end{align}
where $c^\dagger_l$ and $c_l$ are the electron creation and annihilation operators at site $l$, respectively,
$t$ is the hopping between the neighboring edge sites,
$t'$ is that between the two neighboring sites at the corner, $l=\pm1$,
and $t''$ is the second-nearest neighbor hopping between $(l,l') = (2,-1)$ and $(1,-2)$.
By using a similar procedure to Eqs.\ (\ref{H_nf}) and (\ref{H_eff}), we obtain the hopping parameters, $t=0.060$ eV, $t'=0.074$ eV, and $t''=0.114$ eV.

Figure \ref{corner_tb} shows the energy spectrum and the wavefunctions of the corner states obtained from 
(a) the DFT-based full tight binding model, and from (b) the effective edge-site model.
We can see that the corner states are well reproduced by the effective model.
Here it should be noted that the corner states emerge just by connecting $p_{x'}$ and $p_{y'}$ edge modes,
without the aid of the corner-isolated orbitals ($p_z$ for the right corner).

\section{Conclusion}
\label{sec_con}

In this paper, we have investigated edge and corner states in monolayer black phosphorene, 
and shown that the multi-orbital band structure under a non-planer puckered structure 
forces the emergence of the edge states at a boundary along an arbitary crystallographic directions.
There the presence of three $p$ orbitals causes formation of a Wannier orbital at every bond center,
and hence cutting any bonds always results in in-gap states through a half breaking of the Wannier orbital.
At a corner where two edges intersect, we find that unexpected multiple corner states appear
due to unavoidable hybridization of the higher-order topological corner state and the edge states.
These characteristic properties are intuitively understood using a topologically-equivalent, analytically-solvable model
where all the bond angles in the phosphorene are deformed to 90$^\circ$.
We expect that the analysis also applies to different materials having a similar puckered honeycomb lattice, such as GaSe, GaS, SnSe, and PbS
\cite{RevModPhys.93.011001}.

\begin{acknowledgments}
This work was supported by JSPS KAKENHI Grant numbers JP21J20403, JP20K14415, JP20H01840, JP20H00127,
and by JST CREST Grant Number JPMJCR20T3, Japan.
T.K. is partially supported by JSPS Core-to-Core program.
\end{acknowledgments}

\appendix

\section{Details of DFT-based tight-binding model}\label{sec:hop}
\begin{figure}[b]
	\begin{center}
	\leavevmode
	\includegraphics[width=0.8\linewidth]{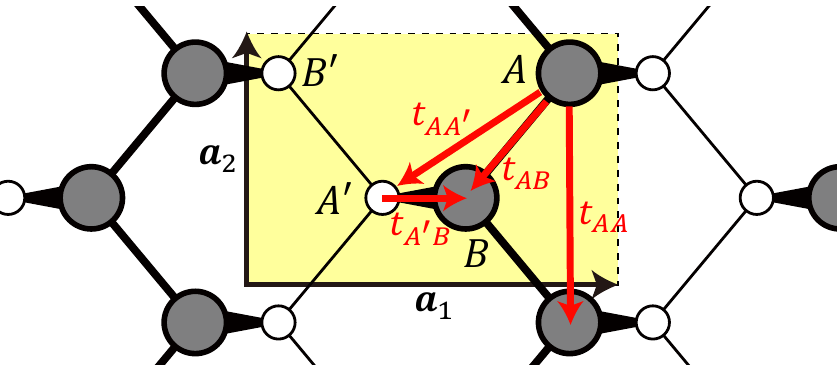}
	\caption{
	Typical nearest hopping between $\mathrm{A'A}$, $\mathrm{BA}$, $\mathrm{AA}$, and $\mathrm{AA'}$ sites. 
	}
	\label{fig:hopping}
	\end{center}
\end{figure}

In this appendix, we present the details of the DFT-based tight binding model obtained from the Wannier90 package~\cite{wannier90}.
In the calculation demonstrated in the main text, we take into account the hopping within the range of distance $6a_1$ ($\sim180$ nm), which strongly decay with the relative distance between the atoms.
Complete sets of the hopping parameters are given in the supplemental data~\cite{sm}.
In the Table~\ref{table:hopping}, we provide the representative parameters for 
onsite potential and hopping for the A site and hopping between nearest A$'$B, AB, AA, and AA$'$ sites [Fig.~\ref{fig:hopping}].
For each pair of the atomic sites, 16 components of hopping parameters describing hopping between the four orbitals are assigned.
Every hopping integral is real valued, due to the time-reversal symmetry.
These values are consistent with the orbital property and geometrical locations.
For instance, for a pair of sites A$'$ and B shown in Fig.\ref{phos_atom}(b), the hopping integral between $p_z$ orbitals is strongest with $t\sim 2.5$ eV, 
since these sites are almost vertically aligned in $z$ direction and $p_z$ orbitals form the $\sigma$ bonds.
The hopping integrals for AB$'$, A$'$B$'$, A$'$A$'$, BB, B$'$B$'$ and BB$'$ and onsite potential and hopping for B$'$, A$'$, and B sites are generated from the Table~\ref{table:hopping} by symmetry operations.

\begin{table}[b]
	\caption{Typical onsite potential and hopping integral $\left<\bm{R},S_2,m_2|H|\bm{0},S_1,m_1\right>$ in the unit of eV,
	where $|\bm{R},S,m\rangle$ indicates the basis of the orbital $m=s, p_x, p_y, p_z$
	at the sublattice $S=A, B', A', B$ in the unit cell located at $\bm{R}$.
	The unit cell is defined as in Fig.~\ref{fig:hopping}.
	Labels of column (row) in the table stand for the orbital $m_1$ ($m_2$).
	Onsite potential and hopping is for the $A$-site.
	The hopping parameter between nearest A$'$ and B, A and B, A and A$'$ with $\bm{R}=\bm{0}$,
	and between A and A with $\bm{R}=-\bm{a}_2$ are presented.
	} \label{table:hopping}
	\renewcommand{\arraystretch}{1.5}
	{\tabcolsep = 2.0mm
		\begin{tabular*}{85mm}{c|cccc}
		\hline 
		\hline 
		\!\!Onsite\!\! & $s$ & $p_x$ & $p_y$ & $p_z$ \\ \hline
		$s$ & $-11.687197$ & $-0.040782$ & $0$ & $0.036646$ \\
		$p_x$ & $-0.040782$ & $-4.225991$ & $0$ & $-0.125509$ \\
		$p_y$ & $0$ & $0$ & $-4.140375$ & $0$ \\
		$p_z$ & $0.036646$ & $-0.125509$ & $0$ & $-4.256935$ \\
		\hline 
		\hline
	\end{tabular*}
	}
	\renewcommand{\arraystretch}{1}

	\vspace{3mm}
	
	\renewcommand{\arraystretch}{1.5}
	{\tabcolsep = 2.3mm
	\begin{tabular*}{85mm}{c|cccc}
		\hline 
		\hline 
		$t_{\mathrm{A'B}}$ & $s$ & $p_x$ & $p_y$ & $p_z$ \\ \hline
		$s$ & $-1.690771$ & $-0.816721$ & $0$ & $-2.437822$ \\
		$p_x$ & $0.816721$ & $-0.536829$ & $0$ & $1.325923$ \\
		$p_y$ & $0$ & $0$ & $-1.10449$ & $0$ \\
		$p_z$ & $2.437822$ & $1.325923$ & $0$ & $2.507026$ \\

		\hline 
		\hline
	\end{tabular*}

	\vspace{3mm}

	\begin{tabular*}{85mm}{c|cccc}
		\hline 
		\hline 
		$t_{\mathrm{AB}}$ & $s$ & $p_x$ & $p_y$ & $p_z$ \\ \hline
		$s$ & $-1.778623$ & $1.827083$ & $1.953456$ & $-0.008235$ \\
		$p_x$ & $-1.827083$ & $0.683126$ & $1.977488$ & $0.000059$ \\
		$p_y$ & $-1.953456$ & $1.977488$ & $1.344737$ & $0.016749$ \\
		$p_z$ & $-0.008235$ & $-0.000059$ & $-0.016749$ & $-1.179879$ \\
		\hline 
		\hline
	\end{tabular*}

	\vspace{3mm}

	\begin{tabular*}{85mm}{c|cccc}
		\hline 
		\hline 
		$t_{\mathrm{AA}}$ & $s$ & $p_x$ & $p_y$ & $p_z$ \\ \hline
		$s$ & $0.030988$ & $-0.078703$ & $0.116371$ & $-0.01095$ \\
		$p_x$ & $-0.078703$ & $-0.319182$ & $0.306764$ & $-0.019294$ \\
		$p_y$ & $-0.116371$ & $-0.306764$ & $0.598276$ & $0.006828$ \\
		$p_z$ & $-0.01095$ & $-0.019294$ & $-0.006828$ & $-0.006613$ \\
		\hline 
		\hline
	\end{tabular*}

	\vspace{3mm}
	\begin{tabular*}{85mm}{c|cccc}
		\hline 
		\hline 
		$t_{\mathrm{AA'}}$ & $s$ & $p_x$ & $p_y$ & $p_z$ \\ \hline
		$s$ & $0.008801$ & $0.021139$ & $0.004026$ & $-0.072713$ \\
		$p_x$ & $0.060405$ & $0.165221$ & $0.061229$ & $0.107306$ \\
		$p_y$ & $-0.039265$ & $-0.096533$ & $-0.03637$ & $-0.318646$ \\
		$p_z$ & $-0.046967$ & $0.074745$ & $-0.092788$ & $-0.091795$ \\
		\hline 
		\hline
	\end{tabular*}
	}
	\renewcommand{\arraystretch}{1}
\end{table}

\section{Symmetry representation and Wannier center for black phosphorene}\label{sec:sym}

As discussed in the main text, the central position of the Wannier orbital
plays a key role in the emergence of the edge and corner states of black phosphorene.
In Sec.~\ref{sub_90}, we identified these positions by using an effective $90^\circ$ model which is topologically equivalent to phosphorene. 
On the other hand, there is an alternative generic scheme to obtain the Wannier center
by using the spacial symmetry and the irreducible representation~\cite{PhysRevX.7.041069,Song2017, Po2017, Bradlyn2017, Cano2021}.
In this appendix, we apply the latter scheme to black phosphorene with space group $Pmna$,
and obtain the consistent result with the main text.

	\begin{figure}
		\begin{center}
		\leavevmode
		\includegraphics[width=80mm]{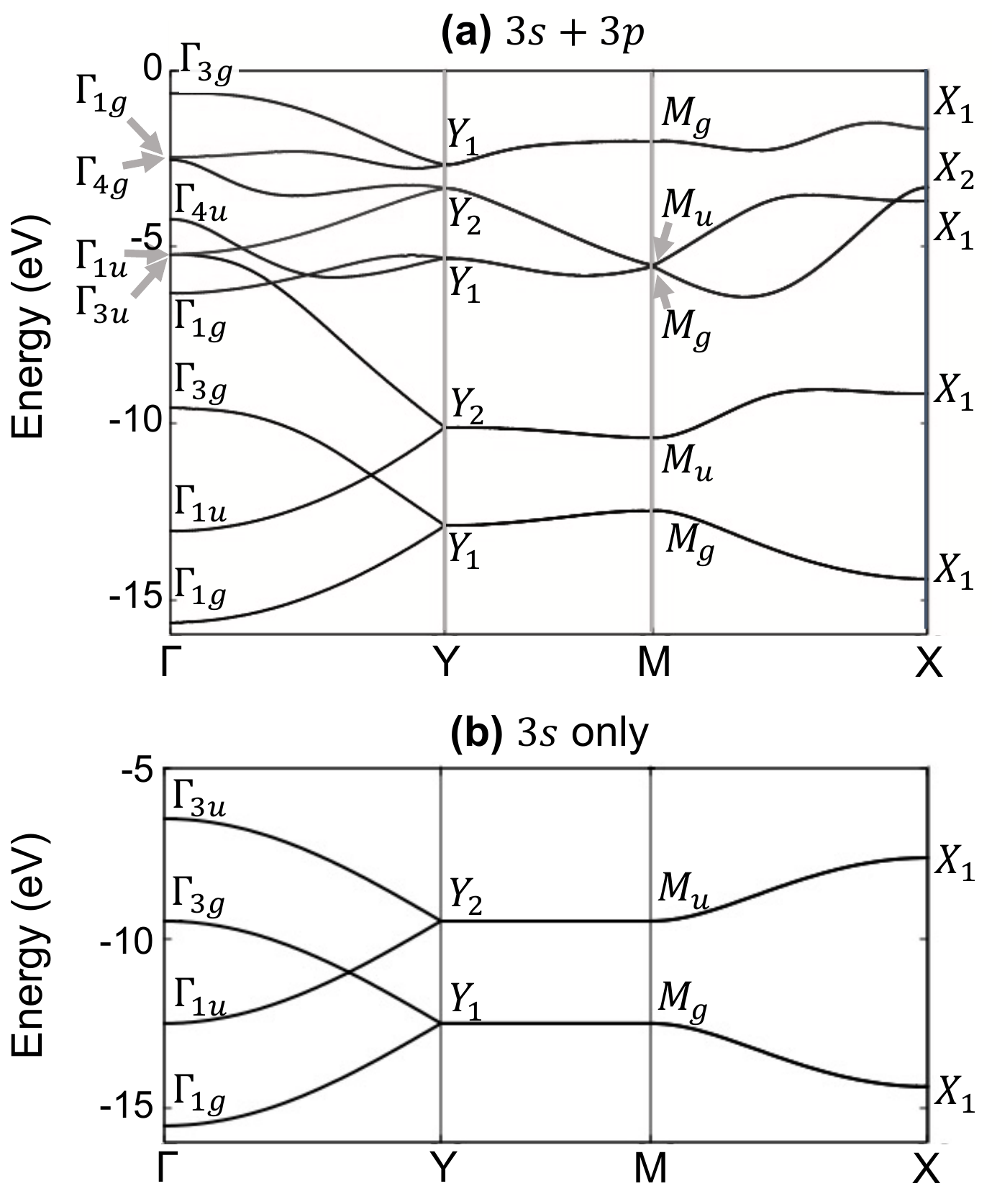}
		\caption{
		(a) Irreducible representation of occupied bands in the DFT-based tight-binding model of phosphorene, which include $3p$ and $3s$ orbital
		[corresponding to the $E<0$ region of Fig.~\ref{band_0_to_1}(b)]. $E=0$ is the
		Fermi energy.
		(b) Irreducible representation of the $3s$-orbital-only model [$E<-5$ eV region of Fig.~\ref{band_0_to_1}(c)].
		}
		\label{band_irrep}
		\end{center}
	\end{figure}

\begin{table}[tb]
	\caption{Irreducible representation of space group $Pmna$.
	$\Gamma_{js}$, $X_i$, $Y_i$, and $M_s$ are irreps at $\Gamma$, $X$, $Y$, $M$ points in the Brillouin zone, respectively.
	Symmetry index is listed from 2nd to 5th columns. 
	$C_{2y}$ is the 180$^\circ$ rotation along the $y$ axis, 
	$M_y$ is the mirror reflection with respect to the $xz$ plane,
	$\{M_z|\tfrac{1}{2}\tfrac{1}{2}\}$ is the glide mirror reflection, i.e., the combination of half translation 
	$\bm{t}=(\bm{a}_1+\bm{a}_2)/2$ and mirror reflection with respect to $xy$ plane, and 
	$P$ is inversion.
	} \label{table:irrep}
	\renewcommand{\arraystretch}{1.5}
	{\tabcolsep = 2.3mm
	\begin{tabular*}{60mm}{c|cccc}
		\hline 
		\hline 
		Irrep & $C_{2y}$ & $M_y$ & $\{M_z|\tfrac{1}{2}\tfrac{1}{2}\}$ & $P$ \\ \hline
		 $\Gamma_{1g}$  & $+1$  & $+1$  & $+1$  & $+1$   \\
		 $\Gamma_{2g}$  & $-1$  & $-1$  & $-1$  & $+1$   \\
		 $\Gamma_{3g}$  & $+1$  & $+1$  & $-1$  & $+1$   \\
		 $\Gamma_{4g}$  & $-1$  & $-1$  & $+1$  & $+1$   \\
		 $\Gamma_{1u}$  & $-1$  & $+1$  & $-1$  & $-1$   \\
		 $\Gamma_{2u}$  & $+1$  & $-1$  & $+1$  & $-1$   \\
		 $\Gamma_{3u}$  & $-1$  & $+1$  & $+1$  & $-1$   \\
		 $\Gamma_{4u}$  & $+1$  & $-1$  & $-1$  & $-1$   \\  
		\hline 
		 $X_{1}$  & 0  & $+2$  & 0  & 0   \\  
		 $X_{2}$  & 0  & $-2$  & 0  & 0   \\  
		\hline 
		 $Y_{1}$  & $+2$  & 0  & 0  & 0   \\  
		 $Y_{2}$  & $-2$  & 0  & 0  & 0   \\  
		\hline 
		 $M_{g}$  & 0  & 0  & 0  & $+2$   \\  
		 $M_{u}$  & 0  & 0  & 0  & $-2$   \\  
		\hline 
		\hline
	\end{tabular*}
	}
	\renewcommand{\arraystretch}{1}
\end{table}

In general, a set of bands is characterized by the irreducible representations (irreps) at high symmetry momenta.
In the space group $Pmna$, specifically, we have irreps at $\Gamma$, $X$, $Y$, and $M$ points summarized in Table~\ref{table:irrep}. 
By using these irreps, we can describe character of the occupied bands below the gap by a single vector
\begin{eqnarray}\label{eq:bvec}
	\bm{b}=(\gamma_{1g},\cdots,\gamma_{4g};\gamma_{1u},\cdots,\gamma_{4u};\xi_{1},\xi_{2};\eta_{1},\eta_{2};\mu_{g},\mu_{u}), 
\end{eqnarray}
where $\gamma_{js}$, $\xi_{j}$, $\eta_{j}$, and $\mu_s$ ($j=1,2,\cdots$ and $s=u$, $g$) is the number of irreps $\Gamma_{js}$, $X_i$, $Y_i$, and $M_s$, respectively, in the occupied bands.
Considering the symmetry of the wavefunctions obtained from the DFT-based tight-binding model in Sec.~\ref{sub_tb}, 
we identify the irreps of the occupied bands as shown in Fig.~\ref{band_irrep}(a).
The vector Eq.~(\ref{eq:bvec}) for the occupied bands in the cluster of $3s$+$3p$ orbital is
\begin{eqnarray}\label{eq:bbp}
	\bm{b}_{3s+3p}=(3,0,2,1;2,0,1,1;8,2;6,4;6,4).
\end{eqnarray}
It is clear that $3s$ bands are located far below the Fermi energy,
and hence we can separate $3s$ bands from $3p$ bands
by continuously shifting the $3s$ cluster to lower energy without topological change (gap closing) at the Fermi energy.
The band representation of $3p$ bands, which of our interest, can be obtained just by subtracting the irreps of $3s$ bands from $3s+3p$.
The irreps of $3s$  can be obtained by an arbitrary tight-binding model having only $s$-orbitals at phosphorus sites.
For example, Figure \ref{band_irrep}(b) shows the irreps of $s$-orbital sector of 90$^\circ$ model [Eq.~(\ref{H90})],
giving  $\bm{b}_{3s}=(1,0,1,0;1,0,1,0;4,0;2,2;2,2)$. 
Therefore, the band representation of remaining $3p$ bands is
\begin{eqnarray}
	\bm{b}_{3p} = \bm{b}_{3s+3p} - \bm{b}_{3s} 
	= (2,0,1,1;1,0,0,1;4,2;4,2;4,2).
\end{eqnarray}

The numbers $\gamma_{js}$, $\xi_{j}$, $\eta_{j}$, and $\mu_s$ in Eq.~(\ref{eq:bvec}) are not independent but
related by the following compatibility conditions,
\begin{eqnarray}
	& \gamma_{1g}\!+\!\gamma_{2u}\!+\!\gamma_{3u}\!+\!\gamma_{4g}\! =\! \gamma_{1u}\! +\! \gamma_{2g}\! +\! \gamma_{3g}\! +\! \gamma_{4u}\! \equiv\! M, &\label{eq:cr1} \\
	& \xi_{1} = \gamma_{1g} + \gamma_{3g} + \gamma_{1u} + \gamma_{3u}, \label{eq:cr2}\\
	& \eta_{1} = \gamma_{1g} + \gamma_{2u} + \gamma_{3g} + \gamma_{4u}, \label{eq:cr3}\\
	& \xi_{1} + \xi_{2} = 2M, \label{eq:cr4}\\
	& \eta_{1} + \eta_{2} = 2M , \label{eq:cr5}\\
	& \mu_g + \mu_{u} = 2M \label{eq:cr6},
\end{eqnarray}
which guarantee the existence of the band gap between the occupied and unoccupied bands.   
In fact, conditions Eq.~(\ref{eq:cr1})-(\ref{eq:cr3}) forbid the band crossing of opposite-parity states 
under the glide mirror reflection $\{M_z|\tfrac{1}{2}\tfrac{1}{2}\}$, 
mirror reflection $M_{y}$, and twofold rotation $C_{2y}$, respectively. 
In addition, the conditions Eqs.~(\ref{eq:cr4})-(\ref{eq:cr6}) make the total number of occupied bands constant $2M$ for each $\bm{k}$ points. 
The conditions Eq.~(\ref{eq:cr1})-(\ref{eq:cr6}) reduce the 14 component degrees of freedom for the vector in Eq.~(\ref{eq:bvec}) to 8 component, 
\begin{eqnarray}\label{eq:bred}
	\tilde{\bm{b}}=(\gamma_{1g},\gamma_{2g},\gamma_{3g},\gamma_{4g};\gamma_{1u},\gamma_{2u},\gamma_{3u};\mu_{g})
\end{eqnarray}
The specific value for the black phosphorene is 
\begin{eqnarray}\label{eq:bredbp}
	\tilde{\bm{b}}_{3p}=(2,0,1,1;1,0,0;4).
\end{eqnarray}

\begin{figure}[t]
	\begin{center}
	\includegraphics[width=85mm]{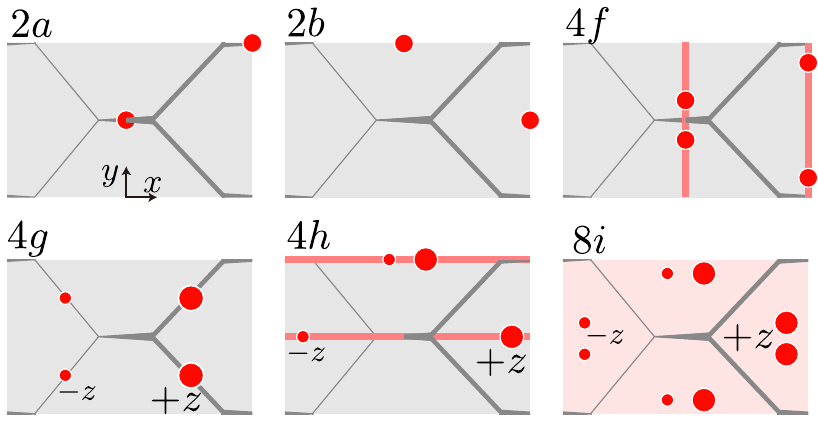}
	\caption{
	Wyckoff position (red dots) of space group $Pmna$. 
	Position $4f$ and $4h$ can move on red thick lines,
	and $8i$ is general position.
	Large and small dots (in $4g, 4h, 8i$) indicate the height of $+z$ and $-z$, respectively, and the middle dots (in $2a, 2b, 4f$) are at $z=0$. 
	Gray rectangle is a unit cell, and gray lines represent the atomic bonds of phosphorene.}
	\label{fig:wyckoff}
	\end{center}
\end{figure}

As a next step, we list the band character Eq.~(\ref{eq:bred}) for all possible elementary bands allowed in the space group $Pmna$.
The elementary band is the band structure obtained from a possible arrangement of atomic orbitals located at a Wyckoff position (WP).
Figure \ref{fig:wyckoff} summarizes all the possible WPs  for $Pmna$.
The WPs are classified by site symmetry group (SSG), or the symmetry group which keeps the WP invariant.
For example, the characterisitic SSG for the WP 2a is $C_{2h}$, which is generated 
by inversion about the unit cell center (the center of the gray rectangle in Fig.~\ref{fig:wyckoff}) 
and two fold rotation around $y$ axis. The SSG for each WP are presented in the second column of the Table~\ref{table:EBindex}. 

\begin{table}[t]
	\caption{Elementary band representations for the space group $Pmna$.
	The low with $\chi$ is the label of representations. 
	The index of Eq.~(\ref{eq:EBindex}) are listed from 5th-12th column.
	} \label{table:EBindex}
	\renewcommand{\arraystretch}{1.5}
	{\tabcolsep = 1.mm
	\begin{tabular*}{85mm}{ccc|c|cccccccc}
		\hline 
		\hline 
		WP & SSG & irreps  & $\chi$&  $\gamma_{1g}$ & $\gamma_{2g}$ & $\gamma_{3g}$& $\gamma_{4g}$ & $\gamma_{1u}$ & $\gamma_{2u}$& $\gamma_{3u}$& $\mu_{g}$ \\ \hline
		2a & $C_{2h}$ & $A_{g}$ &1& 1 & 0 & 1 & 0 & 0 & 0 & 0 & 2  \\
		   &          & $A_{u}$ &2& 0 & 0 & 0 & 0 & 1 & 0 & 1 & 0  \\
		   &          & $B_{g}$ &3& 0 & 1 & 0 & 1 & 0 & 0 & 0 & 2  \\
		   &          & $B_{u}$ &4& 0 & 0 & 0 & 0 & 0 & 1 & 0 & 0  \\ \hline
		2b & $C_{2h}$ & $A_{g}$ &5& 1 & 0 & 1 & 0 & 0 & 0 & 0 & 0  \\
		   &          & $A_{u}$ &6& 0 & 0 & 0 & 0 & 1 & 0 & 1 & 2  \\
		   &          & $B_{g}$ &7& 0 & 1 & 0 & 1 & 0 & 0 & 0 & 0  \\
		   &          & $B_{g}$ &8& 0 & 0 & 0 & 0 & 0 & 1 & 0 & 2  \\ \hline 
		4e & $C_{1h}$ & $A$     &9& 1 & 0 & 1 & 0 & 0 & 1 & 0 & 2  \\
		   &          & $B$     &10& 0 & 1 & 0 & 1 & 1 & 0 & 1 & 2  \\ \hline 
		4g & $C_{2z}$ & $A$     &11& 1 & 0 & 0 & 1 & 1 & 0 & 0 & 2  \\
		   &          & $B$     &12& 0 & 1 & 1 & 0 & 0 & 1 & 1 & 2  \\ \hline 
		4h & $C_{2y}$ & $A$     &13& 1 & 0 & 1 & 0 & 1 & 0 & 1 & 2  \\
		   &          & $B$     &14& 0 & 1 & 0 & 1 & 0 & 1 & 0 & 2  \\ \hline 
		8i & $C_1$    & $A$     &15& 1 & 1 & 1 & 1 & 1 & 1 & 1 & 4  \\ 
		\hline 
		\hline
	\end{tabular*}
	}
	\renewcommand{\arraystretch}{1}
\end{table}

An array of atomic orbitals at WP should be an irrep of the corresponding SSG,
where different symmetries of the atomic orbital (e.g., $s$-like, $p_x$-like) give different irreps. 
In Table~\ref{table:EBindex}, we list all the possible irreps for each WP, and label them with a serial number $\chi=1$ to 15.
In the same manner as Eq.~(\ref{eq:bred}), elementary bands of $\chi$ are described as 
\begin{eqnarray}\label{eq:EBindex}
	\tilde{\bm{b}}_{\chi}=(\gamma_{1g}^{(\chi)},\gamma_{2g}^{(\chi)},\gamma_{3g}^{(\chi)},\gamma_{4g}^{(\chi)};\gamma_{1u}^{(\chi)},\gamma_{2u}^{(\chi)},\gamma_{3u}^{(\chi)};\mu_{g}^{(\chi)}),
\end{eqnarray}
which are listed in the Table~\ref{table:EBindex}.
Actually, the vector $\tilde{\bm{b}}$ in Eq.~(\ref{eq:bred}) for any atomic insulator are always decomposed to a linear combination of $\bm{\bm{b}}_{\chi}$.
The Wannier orbitals and their center position of the system can be obtained by such a decomposition.

In phosphorene, the decomposition into the elementary bands is written as
\begin{eqnarray}\label{eq:bprel}
	\tilde{\bm{b}}_{3p} = \sum_{\chi} n_{\chi} \tilde{\bm{b}}_{\chi},
\end{eqnarray}
where $\tilde{\bm{b}}_{\mathrm{3p}}$ is given by Eq.~ (\ref{eq:bredbp}), and $n_{\chi}$ must be 0 or a positive integer.
Here we have a unique solution, 
\begin{eqnarray}
	n_{\chi} = 
	\left\{ \begin{array}{ll}
	1  &(\chi= 1, 11),\\
	0  & \rm{(otherwise)}. 
	\end{array}\right.
\end{eqnarray}
Because $\bm{b}_{1}$ and $\bm{b}_{11}$ are $s$-like orbitals at the Wyckoff positions $1a$ and $4g$ respectively,
we conclude that the Wannier centers for the phosphorene are located at the every midpoint of the nearest neighboring atoms.
This is consistent with the results from the 90$^\circ$ in the main text.


\bibliography{phosedge}
\end{document}